\documentclass[aps,prd,reprint,longbibliography]{revtex4-2}

\usepackage{graphicx}
\usepackage{amsmath,amssymb}
\usepackage{bm}
\usepackage{xcolor}
\usepackage{hyperref}
\usepackage{orcidlink}

\newcommand{\bT}{\beta}
\newcommand{\aB}{\alpha_B}

\begin{document}

\title{Probing Scalar Polarization with Next-Generation Gravitational-Wave Detectors: The Roles of Detector Geometry and Source Inclination}

\author{Sk Md Adil Imam\orcidlink{0000-0003-3308-2615}}
 \email{adil.imam@unab.cl}
  \affiliation{Institute of Astrophysics, Department of Physics and Astronomy, Universidad Andrés Bello, Santiago, Chile}
   \author{Macarena Lagos\orcidlink{0000-0003-0234-9970}}
 \email{macarena.lagos.u@unab.cl}
  \affiliation{Institute of Astrophysics, Department of Physics and Astronomy, Universidad Andrés Bello, Santiago, Chile}
  
\date{\today}

\begin{abstract}
Gravitational-wave observations of compact binaries offer a direct way to test
whether gravity admits polarizations beyond the two tensor modes of general
relativity. We perform a forecast to assess how well future multi-messenger events will constrain the presence of scalar polarizations in gravitational waves. Performing binary
neutron star injections, we forecast next-generation constraints on three 
modified gravity parameters in the PPE formalism: $\alpha$ (tensor amplitude), $\aB$ (scalar amplitude) and $\bT$ (tensor and scalar phase). Cosmic Explorer (CE) and Einstein Telescope (ET) each
tighten every parameter by 1--2 orders of magnitude over an O5 LIGO-Virgo network, with 
ET's triangular geometry giving a persistent $\approx2\times$ improvement over 2$\times$CE on the scalar-mode
amplitude $\alpha_B$ in both a GW170817-like and a second,
representative BNS.
We show that, if the sky position and luminosity distance are known from EM observations, adding an EM-based prior on the orbital inclination $\iota$ narrows the $\alpha_B$ constraint by $\approx56\times$ relative to an isotropic inclination prior, for a GW170817-like event.
We also assess how EM inclination priors as well as assumptions on the neutron star equation of state may affect the inference on the scalar amplitude $\alpha_B$.
Overall, these
results indicate that detector geometry and the availability of
well-placed EM external information are both important
levers --- alongside raw network sensitivity --- for next decade
scalar-polarization tests.
\end{abstract}
\maketitle

\section{Introduction}
\label{sec:intro}

While General Relativity (GR) predicts only two transverse tensor
polarization modes for gravitational waves, generic metric theories of
gravity allow up to six independent polarization
modes~\cite{Weinberg1972,MTW1973,Eardley_PRD,Will_1993,Maggiore2008,Schutz2009}. Scalar-tensor
theories generically predict an additional scalar breathing
mode~\cite{Will:2014kxa}, while more general classes of
modified gravity --- vector-tensor~\cite{Will:2014kxa}, tensor-vector-scalar \cite{Bekenstein_2004,Bekenstein_2006}, Einstein-aether~\cite{Jacobson_2004,eling2005,jacobson2008},
and bimetric theories~\cite{Will:2014kxa,Chatziioannou:2012rf,Lightman_73,ROSEN_1974} among them --- admit up to the full complement of
non-tensorial polarizations. Detecting, or meaningfully bounding, any
non-tensorial polarization content in a gravitational-wave signal would
constitute direct evidence that GR is incomplete, independent of any
assumption about which specific alternative theory is responsible.

Binary neutron star (BNS) mergers offer a distinct advantage for such
tests: unlike binary black holes, they can produce electromagnetic (EM)
counterparts whose independently measured sky position, distance,
orbital inclination  and orbital orientation, break parameter degeneracies that otherwise limit
gravitational-wave-only polarization tests. GW170817, the first
confirmed multi-messenger BNS event, remains the only source for which
this combination has been exploited in
practice~\cite{LIGOScientific:2017zic}. In a previous analysis, we used
the angular information from GW170817's electromagnetic
counterparts --- including, for the first time, the polarization-angle
constraint from the gamma-ray burst afterglow --- to perform a
parameterized post-Einsteinian (PPE)~\cite{Yunes_2009,Chatziioannou:2012rf,Tahura:2018zuq} test of GR incorporating both
tensor deformations and a scalar breathing
polarization~\cite{Imam_2026}. That analysis found a mild, $2$--$3\sigma$
preference for a non-zero scalar amplitude, though Bayesian model
comparison still favored pure GR once the PPE model's additional
parameters were penalized. We obtained  $\Delta\log\mathcal{Z} =
4.7\pm0.3$ for the log-Bayes factor, corresponding to strong evidence for GR on Jeffreys' scale.

This raises an immediate question: how much would that constraint
improve with next-generation gravitational-wave detectors? Cosmic
Explorer (CE)~\cite{Reitze:2019iox} and the Einstein Telescope (ET)~\cite{Punturo:2010zz} are each expected to
improve strain sensitivity by roughly an order of magnitude over the
current LIGO-Virgo-KAGRA network, with implications for tests of GR
that have been explored broadly across the modified-gravity
literature~\cite{Chamberlain:2017fjl,Perkins:2020tra,ET:2019dnz}. Forecasts using the PPE and
related frameworks indicate that next-generation, multi-band, and
multi-detector observations can tighten bounds on generic non-GR
parameters by one to several orders of
magnitude~\cite{Perkins:2020tra,Carson:2019rda}, and
detection-capability forecasts built directly on the CE and ET design
sensitivities used here confirm the scale of this
improvement~\cite{Iacovelli:2022bbs,IacovelliGWFAST2022,Srivastava2022CosmicExplorerDesign,Branchesi2023ETDesigns}.
Specifically for polarization content, in~\cite{Takeda:2019gwk} the authors showed
that next-generation ground-based networks --- including configurations
resembling CE and ET --- substantially improve the prospects for
distinguishing tensor, vector, and scalar polarizations from compact
binary mergers relative to current-generation
detectors. What remains largely
unexplored is how much of that improvement comes from raw detector
sensitivity and network geometry alone, versus how much depends on
what external information --- particularly on the source's orbital
inclination or orientation --- is available to break the degeneracy between the breathing-mode scalar amplitude and orbital parameters that may limit GW-only
analyses.

Following the same framework as our previous work on GW170817 \cite{Imam_2026}, in this paper we forecast next-generation constraints on three PPE parameters:
tensor modified amplitude parameter ($\alpha$), tensor and scalar modified phase parameter ($\beta$), and
breathing scalar amplitude parameter ($\alpha_B$). We perform injections for a GW170817-like event as well as a future representative BNS event, for the O5 LIGO-Virgo network, a single Einstein Telescope, and a two-site Cosmic Explorer
network.
We quantify how much of the resulting precision on $\alpha_B$
is driven by detector network geometry and sensitivity, as well as
the information from electromagnetic counterparts. 

This paper is structured as follows:
Sec.~\ref{sec:waveform} summarizes the waveform model and analysis
pipeline. Sec.~\ref{sec:results} presents our five main results ---
network sensitivity, EM-information content, inclination-prior bias, equation of state (EOS) sensitivity for a GW170817-like event and a repeat of the network comparison for a future representative BNS system. Finally,  Sec.~\ref{sec:conclusion} discusses the implications for next-generation scalar polarization tests.

\section{Waveform Model and Analysis Pipeline}

\subsection{Waveform Model}
\label{sec:waveform}

We adopt the same parameterized post-Einsteinian (PPE) waveform model
developed and validated in our companion analysis~\cite{Imam_2026}, to which we refer the reader for more details. Here, we summarize the elements needed to
interpret the results below.

The PPE framework was originally introduced to parameterize
leading-order, theory-agnostic deviations from GR in the tensor sector
of the gravitational-wave phase and amplitude~\cite{Yunes_2009},
and was subsequently extended to the complete polarization content ---
including the scalar breathing mode~\cite{Chatziioannou:2012rf, Akama:2026igu}.
Three free PPE amplitudes are introduced on top of the GR waveform, each
multiplying a fixed frequency-dependent factor $u_m = (2\pi M_z
f/|m|)^{1/3}$, where $M_z$ is the redshifted total mass and $|m|$
selects the leading quadrupole ($\ell=|m|=2$) harmonic.
The quadrupole tensor polarizations are deformed multiplicatively in amplitude and
additively in phase, with respect to GR, in frequency domain
\begin{equation}
\tilde{h}^{(2,2)}_{\rm T}(f) = \tilde{h}^{(2,2)}_{\rm GR,T}(f)
\,(1+\alpha u_2^{a_T})\,e^{i\beta u_2^{b}},
\end{equation}
with $\alpha$ and $\beta$ describing deviations in the tensor amplitude and phase, respectively. In addition, the quadrupole scalar breathing polarization is modeled as
\begin{equation}
    \tilde{h}_b=Y_b(\iota)\,  \tilde{h}^{(2,2)}_{\rm GR,T}(f)
\alpha_B\, u_2^{a_S}e^{i\beta u_2^{b}},
\end{equation}
where $\alpha_B$ is a parameter that determines the amplitude of this scalar polarization,
 and $Y_b(\iota) = \sqrt{3/2}\,\sin^2\iota$ is the
inclination-dependent antenna factor~\cite{Akama:2026igu}
responsible for the leading-order degeneracy between $\alpha_B$ and $\iota$, central to Sec.~\ref{sec:results}. Here we restrict the scalar sector to the quadrupole
($\ell=|m|=2$) breathing mode; we do not consider the
dipole ($\ell=|m|=1$) case, which our previous analysis of
GW170817 found to be disfavored by the
data~\cite{Imam_2026}.

Following~\cite{Imam_2026}, the exponents $(a_T, b, a_S)$ determine the frequency power of the deviations from GR, and are fixed
at a single benchmark point rather than sampled or marginalized over;
throughout this work we use the quadrupole ``Set 1'' point
$(a_T,b,a_S)=(-2,-7,-2)$, the same choice for
which our previous analysis of GW170817 found a mild
$2$--$3\sigma$ preference for a non-zero scalar
amplitude~\cite{Imam_2026}. In the standard PPE PN-counting convention ---
where GR's leading (quadrupole) phase term scales as $u^{-5}$ and a
phase deformation at relative PN order $n$ scales as $u^{2n-5}$ --- the
phase exponent $b=-7$ corresponds to a $-1$PN order correction, the
characteristic scaling of dipolar-radiation departures from GR,
entering one PN order below the leading GR quadrupole term. The
amplitude exponents $a_T=a_S=-2$ are fixed at the same benchmark; since
all three exponents are negative, $u_m^{a_T}$, $u_m^{b}$, and
$u_m^{a_S}$ all grow toward low frequency, so this benchmark
concentrates its signal-to-noise in the early inspiral rather than in
the merger-band information that otherwise dominates recovery of the
underlying GR parameters.

Similar to our previous companion analysis, we perform Bayesian inference on GW170817 restricted to the inspiral, up to the cutoff frequency
$M_z f_c^{(2,2)} = 0.018$ separating the inspiral from the
intermediate/merger regime --- beyond which the PPE
framework is no longer physically motivated --- and
include throughout the dominant quadrupole ($\ell=|m|=2$) harmonic only.
Previously, we used the point-particle
\texttt{IMRPhenomD} waveform to model the GR signal (cross-checked in that work against the
tidal \texttt{IMRPhenomD\_NRTidalv2}), but here we apply the PPE
formalism to zero-noise injection forecasts for next-generation
detectors, using the tidal \texttt{IMRPhenomXAS\_NRTidalv3}
waveform~\cite{Pratten:2020fqn,Abac:2023ujg} for $h^{(2,2)}_{\rm GR,T}$.
We use \texttt{NRTidalv3}~\cite{Abac:2023ujg}, the more recent
tidal correction, since $\Lambda_1,\Lambda_2$ are sampled here rather
than checking post hoc. More importantly, Mezzasoma \textit{et
al.}~\cite{Mezzasoma:2026wme} recently showed that
\texttt{IMRPhenomD}'s numerical-relativity calibration uncertainty can
mimic a non-GR deviation at SNRs as low as $\sim 60$ --- well below the
SNRs reached by the next-generation detectors. We therefore adopt the independently calibrated \texttt{IMRPhenomXAS}
family instead, though a dedicated calibration-uncertainty study
analogous to~\cite{Mezzasoma:2026wme} does not yet exist for it. 

\subsection{Detector networks}

In Table~\ref{tab:networks} we summarize the three detector configurations considered in this paper. All quantities refer to the injected GW170817-like signal. The
network SNR is the quadrature sum of the per-detector
matched-filter SNRs for our (zero-noise) injection, and
the duration is the length of the frequency-domain
analysis segment, chosen to exceed the inspiral chirp
time at $f_{\rm min}$ for each network. For reference, the real GW170817 event was detected with a network SNR of 32.4 in O2 data~\cite{LIGOScientific:2017vwq}.

\begin{table}[h!]
\begin{ruledtabular}
\caption{Detector-network configurations used throughout. The network SNR and the analysis duration correspond to the injected GW170817-like event, not to the real data, for which the network SNR was 32.4~\cite{LIGOScientific:2017vwq}. }
\label{tab:networks}
\begin{tabular}{lccc}
Network &Net.\ SNR & $f_\text{min}$/$f_\text{max}$ (Hz) & Duration (s) \\
\hline
HLV      & 162  & 20/2048 & 256  \\
ET$\Delta$  & 1233 & 5/2048  & 8192 \\
2$\times$CE & 3113 & 5/2048 & 8192 \\
\end{tabular}
\end{ruledtabular}
\end{table}

\begin{figure}[h!]
\centering
\includegraphics[width=0.48\textwidth]{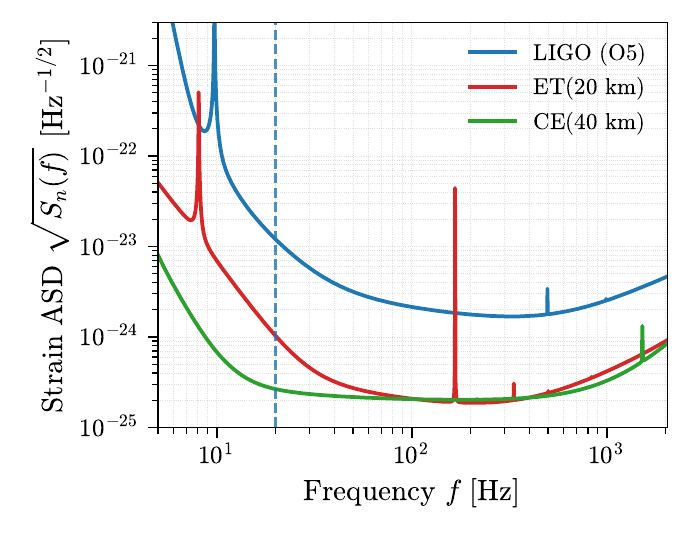}
\caption{Single-interferometer strain amplitude spectral density, $\sqrt{S_n(f)}$, for the three detector designs used throughout this work as in Table~\ref{tab:networks}: the LIGO O5 design curve, the Einstein Telescope (ET, 20\,km) curve, and the Cosmic Explorer (CE, 40\,km) curve. The blue dashed line indicates the minimum frequency of LIGO detector network.} 
\label{fig:asd}
\end{figure}

Fig.~\ref{fig:asd} shows the detectors single-interferometer amplitude spectral densities. Each curve is independent of the number of detectors combined to form a network: For the O5 HLV (Hanford, Livingston, and Virgo) network, we adopt the LIGO A+ design sensitivity targeted for O5~\cite{KAGRA:2013rdx} for each interferometer. For the next-generation networks, we use ET(20\,km) curve for the triangular ET$\Delta$ network, and the CE(40\,km) curve for both sites of the 2$\times$CE network. The blue dashed line indicates the minimum frequency of the HLV detector network single-interferometer amplitude spectral densities. 

ET$\Delta$ and 2$\times$CE~\cite{IacovelliGWFAST2022,Srivastava2022CosmicExplorerDesign,Branchesi2023ETDesigns} are each roughly an order of magnitude quieter
than O5 across the $20$--$500\,$Hz band that dominates the BNS
signal-to-noise ratio (noise floor $\approx2\times10^{-25}\,\mathrm{Hz}^{-1/2}$
for both, near $150$--$200\,$Hz, versus O5's $\approx1.7\times10^{-24}\,
\mathrm{Hz}^{-1/2}$ near $320\,$Hz). The two next-generation curves are not
identical in shape: CE's 40\,km arms give it lower noise below
$\sim100\,$Hz, while ET's underground, cryogenic 20\,km design reaches a
comparably deep low-frequency floor despite its shorter arms, and its
triangular geometry provides three independent, co-located interferometer
outputs from a single site (at the cost of a $60^\circ$ rather than
$90^\circ$ opening angle) instead of the two independent, geographically
separated L-shaped sites that make up CE. For the 2$\times$CE network analyzed here, the two CE interferometers are built with their location and geometry set exactly to the existing LIGO Hanford and Livingston sites, respectively --- the standard one-CE-per-existing-site placeholder convention used in CE forecasting studies. Both next-generation
curves remain low enough below $20\,$Hz to justify the lower
$f_\text{min}=5\,$Hz used for ET$\Delta$ and CE in Table~\ref{tab:networks}, whereas
O5's steeply rising low-frequency noise leaves that band largely
uninformative, motivating its higher $f_\text{min}=20\,$Hz. This geometric
distinction --- one triangular site (ET) versus two independent L-shaped
sites (CE) --- is the physical basis for the network-dependent asymmetry in
$\aB$ sensitivity discussed in Sec.~\ref{sec:result1}.

\subsection{Bayesian inference setup}
\label{sec:bayes}

In this paper, we perform injections of two BNS signals --- a GW170817-like event and a representative BNS taken from \cite{Lagos:2024boe} --- and include mock electromagnetic counterpart information as priors. The likelihood of a  GW signal is modeled as a heterodyned (relative-binning) frequency-domain
Gaussian likelihood, referenced to the injected parameters: the waveform is
evaluated exactly on a coarse grid of frequency bins and linearly
interpolated in between, which reduces the per-sample cost from
$O(n_\text{freq})$ to $O(n_\text{bins})$ without approximating the
likelihood surface near the reference point. This is what keeps the ET$\Delta$ and
2$\times$CE analyses tractable at their $8192\,$s duration and $2048\,$Hz
bandwidth, where the native frequency grid has $\mathcal{O}(10^7)$ bins.

Posteriors are sampled with \texttt{jimgw}~\cite{Edwards:2023sak,Wong:2023lgb,Wouters:2024oxj}, a JAX-native, GPU-accelerated nested sampler, run at $n_\text{live}=2000$ throughout, which simultaneously returns the posterior samples used for every table and figure in this work and the Bayesian evidence $Z$ used as an internal convergence check. All injections are performed in zero (idealized) noise: the
data equal the injected template exactly, so that the recovered posterior
tests the pipeline's precision and bias in isolation, without the
additional realization-to-realization scatter of a single noisy draw.

\begin{table*}[t]
\begin{ruledtabular}
\caption{Prior families and numerical ranges for the 14
sampled parameters used in this analysis, shown here for
the fiducial GW170817-like injection. The Gaussian priors on $d_L$, $\iota$, and
$\psi$ are varied across sections as described in the text;
the values below are the Sec~\ref{sec:result1} baseline.}
\label{tab:priors}
\begin{tabular}{llcc}
Parameter & Symbol & Prior & Injected value \\
\hline
Chirp mass ($M_\odot$) & $M_c$ & Uniform$(1.18,1.21)$ & $1.1975174$ \\
Mass ratio & $q$ & Uniform$(0.5,1.0)$ & $0.6393$ \\
Aligned spins & $\chi_1,\chi_2$ & Uniform$(-0.05,0.05)$ & $0.0283,\ 0.0132$ \\
Luminosity distance (Mpc) & $d_L$ & Gaussian$(42.9,3.2)$ & $42.0$ \\
Inclination (rad) & $\iota$ & Gaussian$(2.8495,0.03)$ & $2.8495$ \\
Polarization angle & $\psi$ & Gaussian$(3.14,0.09)$ & $3.14$ \\
Coalescence phase & $\phi_c$ & Uniform$(0,2\pi)$ & $2.0$ \\
Coalescence time (s) & $t_c$ & Uniform$(-10^{-3},10^{-3})$ & $-2.36\times10^{-4}$ \\
Tidal deformabilities & $\Lambda_1,\Lambda_2$ & Uniform$(0,5000)$ & $791.0,\ 891.0$ \\
Tensor amplitude deformation & $\alpha$ & Uniform$(-10^{-2},10^{-2})$ & $0.0$ \\
Tensor phase deformation & $\beta$ & Uniform$(-10^{-4},10^{-4})$ & $0.0$ \\
Breathing-mode amplitude & $\alpha_B$ & Uniform$(-5,5)$ & $0.0$ \\
\end{tabular}
\end{ruledtabular}
\end{table*}

Each injection is analyzed with a 14-parameter Bayesian model: 11
quasi-circular, aligned-spin, non-precessing GR parameters 
, plus the three PPE deformation parameters $\alpha,\bT,\aB$. Notice that we do not sample the two angles (ra,dec) describing the BNS sky position, since we will assume it to be known through an EM counterpart observation. Motivated by GW170817, whose kilonova counterpart localized it with high precision 
at $(\mathrm{ra},\,\mathrm{dec}) = (13\mathrm{:}09\mathrm{:}48.085 \pm
0.018\,\mathrm{s},\,-23\mathrm{:}22\mathrm{:}53.343 \pm 0.218'')$, host
galaxy NGC~4993~\cite{LIGOScientific:2017zic}, here we fix the sky location at its injected value $(\mathrm{ra},\,\mathrm{dec})=(13\mathrm{:}09\mathrm{:}48.085,-23\mathrm{:}22\mathrm{:}53.343)$. 
All GR-consistent injections use $\alpha=\bT=\aB=0$;
the PPE parameters are always sampled (never fixed), so that the recovered
posterior width around zero is itself the reported bound
(Sec.~\ref{sec:result1}).
For the GR quadrupole ($\ell=|m|=2$) tensor polarizations, we use an unmodified \texttt{IMRPhenomXAS\_NRTidalv3}~\cite{Pratten:2020fqn,Abac:2023ujg})
waveform.
Table \ref{tab:priors} summarizes all of the waveform parameters and their prior ranges considered for our GW170817-like BNS event.

The representative BNS event is taken from \cite{Lagos:2024boe}.
That work places the source at
$\theta_v=30^\circ$ (equivalently $\iota=2.62\,$rad) and assumes an
EM follow-up campaign that measures the inclination to
the same \emph{fractional} precision achieved for GW170817, giving
$\iota=2.62\pm0.06\,$rad, and measures the polarization/jet-orientation
angle from VLBI-resolved afterglow proper motion (a High Sensitivity
Array-like campaign) to $\sigma(\psi)=0.2\,$rad about $\psi=3.14\,$rad. The source is also assumed to be at a distance of  $d_L=100\,$Mpc, with an uncertainty of $\sigma_{d_L}=4.72\,$Mpc which is derived 
following their assumed Euclid/DESI-like spectroscopic precision $\sigma_z=10^{-3}(1+z)$. We convert the redshift measurement (centered at $z_\mu=0.02210$, the value
solved from $D_L(z_\mu)=100\,$Mpc) into a distance uncertainty via the
flat-$\Lambda$CDM distance-redshift relation, propagating both $\sigma_z$
and the Planck 2018 uncertainty on $H_0$~\cite{Planck2018}
($H_0=67.36\pm0.54\,\mathrm{km\,s^{-1}\,Mpc^{-1}}$, $\Omega_{m,0}=0.3153$)
through $\partial D_L/\partial z$ and $\partial D_L/\partial H_0$ in
quadrature (linear error propagation); the redshift term
dominates the resulting budget ($\approx98\%$ of the variance). This gives us 
$d_L\sim\mathcal{N}(100.0,4.72)\,$Mpc, used as a prior for the
representative event. All other parameters --- sky location, chirp mass, mass
ratio, aligned spins, coalescence phase and time, and
tidal deformabilities --- are fixed at the same injected
values as the fiducial GW170817-like source
(Table~\ref{tab:priors}).


\subsection{Statistical conventions}

Precision is reported throughout as the 68\% credible-interval half-width,
$\frac{1}{2}(x_{84}-x_{16})$, computed from the 16th/84th percentiles of the
posterior samples for each parameter, with the median $x_{50}$
quoted as the central value.

Posterior agreement/disagreement between two runs is reported as the
Jensen-Shannon divergence~\cite{LIN_1991},
\begin{equation}
D_{JS}(P\|Q) = \tfrac{1}{2}D_{KL}(P\|M) + \tfrac{1}{2}D_{KL}(Q\|M)
\label{eq:djs}
\end{equation}
with $M = \tfrac{1}{2}(P+Q)$
evaluated in base 2 (bits), so $D_{JS}\in[0,1]$ with $0$ indicating
identical distributions and $1$ indicating fully disjoint support. In
practice, each posterior's Gaussian-KDE density is evaluated on a common
grid spanning the union of the compared runs' samples for that parameter,
normalized to a discrete probability mass function, and
Eq.~\eqref{eq:djs} is computed directly on the two discretized
distributions. Unlike the credible-interval half-width, $D_{JS}$ is
symmetric in $P,Q$ and sensitive to differences in both location and shape,
not just width, so it is used here as a complementary summary statistic
alongside the median/CI values in each table.
\section{Results}
\label{sec:results}

\begin{table*}
\begin{ruledtabular}
\caption{Median and 68\% credible-interval half-width on each PPE parameter, for three GW networks for a GW170817-like event. }
\label{tab:result1}
\begin{tabular}{lccc}
Network & $\alpha$ & $\bT$ & $\aB$ \\
\hline
2$\times$CE  & $(-0.2\pm6.9)\times10^{-6}$ & $(-1.0\pm4.9)\times10^{-10}$ & $(-0.2\pm4.0)\times10^{-4}$ \\
ET$\Delta$  & $(0.0\pm1.4)\times10^{-5}$ & $(-0.7\pm7.1)\times10^{-10}$ & $(0.1\pm1.9)\times10^{-4}$ \\
HLV & $(0.2\pm2.6)\times10^{-4}$ & $(4.7\pm6.7)\times10^{-8}$ & $(0.0\pm1.6)\times10^{-2}$ \\
\end{tabular}
\end{ruledtabular}
\end{table*}

\begin{figure*}
    \centering
    \includegraphics[width=1\textwidth]{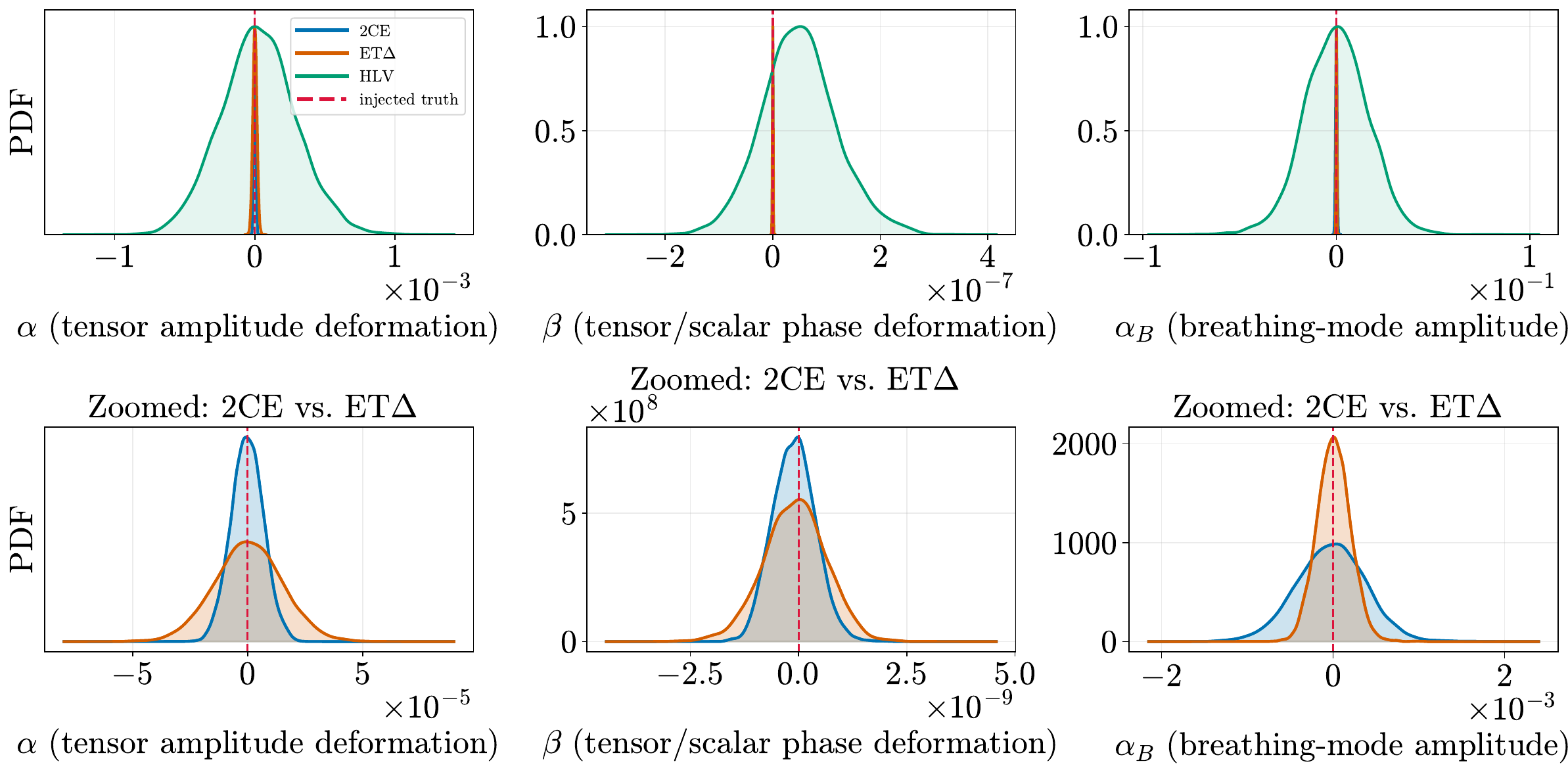}
    \caption{Posterior comparison of $\alpha,\bT,\aB$ across CE, ET, and HLV
(GW170817-like injection), with injected value shown in dashed vertical line. Top row: full range, peak-normalized. Bottom row: zoom to CE vs.\ ET distributions, since
HLV's posterior is $\sim$40--85$\times$ wider on every parameter and would
otherwise be invisible on a shared axis. }
\label{fig:result1}
\end{figure*}

We present five results built on this pipeline, all using GR-consistent
($\alpha=\beta=\alpha_B=0$) binary neutron star injections. Sec.~\ref{sec:result1}
establishes how precisely $\alpha$, $\bT$, and $\aB$ are constrained by a
GW170817-like event across three detector networks: O5 HLV, ET$\Delta$ and $2\times$CE.
Sec.~\ref{sec:result2}
isolates how much of that prior sensitivity comes from different EM observations, by adding informative priors on $\iota$ and then $\psi$
on top of a baseline with known sky position and $d_L$.
Sec.~\ref{Sec:result3} examines how the constraint on $\aB$ shifts when the
assumed prior on the orbital inclination $\iota$ is drawn from different
real external follow-up estimates in the literature~\cite{Hotokezaka2019H0,Ghirlanda2019StructuredJet,Wang2023ViewingAngle}.
Sec.~\ref{sec:eos-sensitivity} tests whether the constraints depend on the
nuclear equation of state, by changing the injected tidal deformabilities.
Finally, Sec.~\ref{sec:result4}
repeats the network comparison of Sec.~\ref{sec:result1} for a second,
independent representative BNS source, testing whether the precision
ordering generalizes beyond a single fiducial event.

\subsection{GW170817: Parameter Precision Across Detector Networks}
\label{sec:result1}

In this section, we mock a GW170817-like event, and assume that the observations of EM counterparts provided a fixed sky location as well as gaussian constraints for $d_L$, $\iota$ and $\psi$ (see Table \ref{tab:priors}). These EM constraints are then used as priors in the Bayesian analysis performed here, where we find the expected constraints on the GW parameters for three detector networks: O5 HLV, ET$\Delta$, and $2\times$CE.

\begin{table*}
\begin{ruledtabular}
\caption{Median and 68\% credible-interval half-width on each PPE
parameter at each stage of the cumulative extrinsic-prior scan, 2$\times$CE
network, sky position fixed at truth throughout. Stage 1: $d_L$ Gaussian
only. Stage 2: $+\iota$ Gaussian. Stage 3: $+\psi$ Gaussian. $\aB$ narrows by $\approx56\times$ from
Stage 1 to 2 and then changes by only $\approx1\times$ from Stage 2
to 3, an end-to-end narrowing of $\approx57\times$ from Stage 1 to
 3 (see Sec.~\ref{sec:result2}). $\alpha$ and $\bT$ are essentially unchanged
across all three stages. }
\label{tab:result3}
\begin{tabular}{llccc}
Stage & EM Priors added & $\alpha$  & $\bT$ & $\aB$ \\
\hline
1 & Sky+$d_L$ & $(0.13\pm7.78)\times10^{-6}$ & $(0.66\pm4.59)\times10^{-10}$ & $(0.02\pm2.25)\times10^{-2}$ \\
2 & $+\,\iota$ & $(0.86\pm8.49)\times10^{-6}$ & $(0.08\pm4.86)\times10^{-10}$ & $(0.14\pm4.04)\times10^{-4}$ \\
3 & $+\,\psi$  & $(-0.15\pm6.94)\times10^{-6}$ & $(-1.01\pm4.90)\times10^{-10}$ & $(-0.18\pm3.95)\times10^{-4}$ \\
\end{tabular}
\end{ruledtabular}
\end{table*}

Fig.~\ref{fig:result1} shows the posterior probability distributions for the three PPE parameters, and Table~\ref{tab:result1} quotes the constrained values. We find that both next-generation networks dramatically outperform O5 HLV on every PPE
parameter, by 1 to 2 orders of magnitude. In particular, for constraining the scalar breathing mode, HLV's 68\% CI on $\aB$ is
$\sim$40--85$\times$ wider than either CE's or ET's constraints, a gap attributable to network sensitivity alone.
Between the two next-generation configurations, the ordering is not
uniform across parameters, although the results are quite comparable. CE is found to provide tighter constraints on $\alpha$ and $\beta$ (by $\approx2\times$ and $\approx1.4\times$, respectively; Table~\ref{tab:result1})), while ET$\Delta$ is $\sim2\times$ tighter on $\aB$
specifically ($\pm1.9\times10^{-4}$ vs.\ CE's $\pm3.9\times10^{-4}$). This asymmetry is consistent with ET's
triangular, three-co-located-detector geometry providing direct
polarization-mode discrimination that our two L-shaped CE
detectors --- oriented nearly in parallel, much like the
LIGO H/L pair --- cannot replicate, even though CE's higher
per-detector strain sensitivity gives it the edge on the
purely tensor-sector parameters $\alpha$ and $\beta$ (see
Fig.~\ref{fig:asd}). 
If confirmed at other source configurations, this would suggest CE and ET$\Delta$ are not simply
interchangeable ``next-generation'' points on a single SNR axis for
scalar-polarization tests specifically --- ET's geometry may be
more valuable for $\aB$ even where its raw network SNR ($\approx1233$ for ET$\Delta$ vs.\ $\approx3113$ for $2\times$CE) is lower than CE's.

It is instructive to compare this idealized O5 HLV forecast to our previous
real-data analysis of GW170817 itself~\cite{Imam_2026}, evaluated at the
same $(a_T, b, a_S) = (-2, -7, -2)$ benchmark and the equivalent EM-informed
prior configuration. That real, noisy-data analysis found
$\alpha = -3.43^{+3.21}_{-3.21} \times 10^{-3}$,
$\beta = 5.77^{+6.65}_{-5.90} \times 10^{-7}$~rad, and
$\alpha_B = -0.29^{+0.18}_{-0.26}$ (95\% CI). At the same 95\% CI, this
idealized zero-noise HLV injection is tighter by a factor of $\approx 4$--$8$
across the three PPE parameters ($\alpha$: $6.1\times$, $\beta$:
$4.2$--$4.7\times$, $\alpha_B$: $5.6$--$8.1\times$). This is consistent with the difference in signal strength: GW170817 was observed in O2 data with a network SNR of $\sim$ 32~\cite{LIGOScientific:2017vwq}, whereas the O5 HLV injection has SNR $\sim$ 162 (Table I) at the same distance, a ratio of $\sim$ 5, and statistical uncertainties scale approximately as 1/SNR. 


\subsection{GW170817: EM Counterpart Information}
\label{sec:result2}

\begin{figure*}
    \centering
    \includegraphics[width=0.8\textwidth]{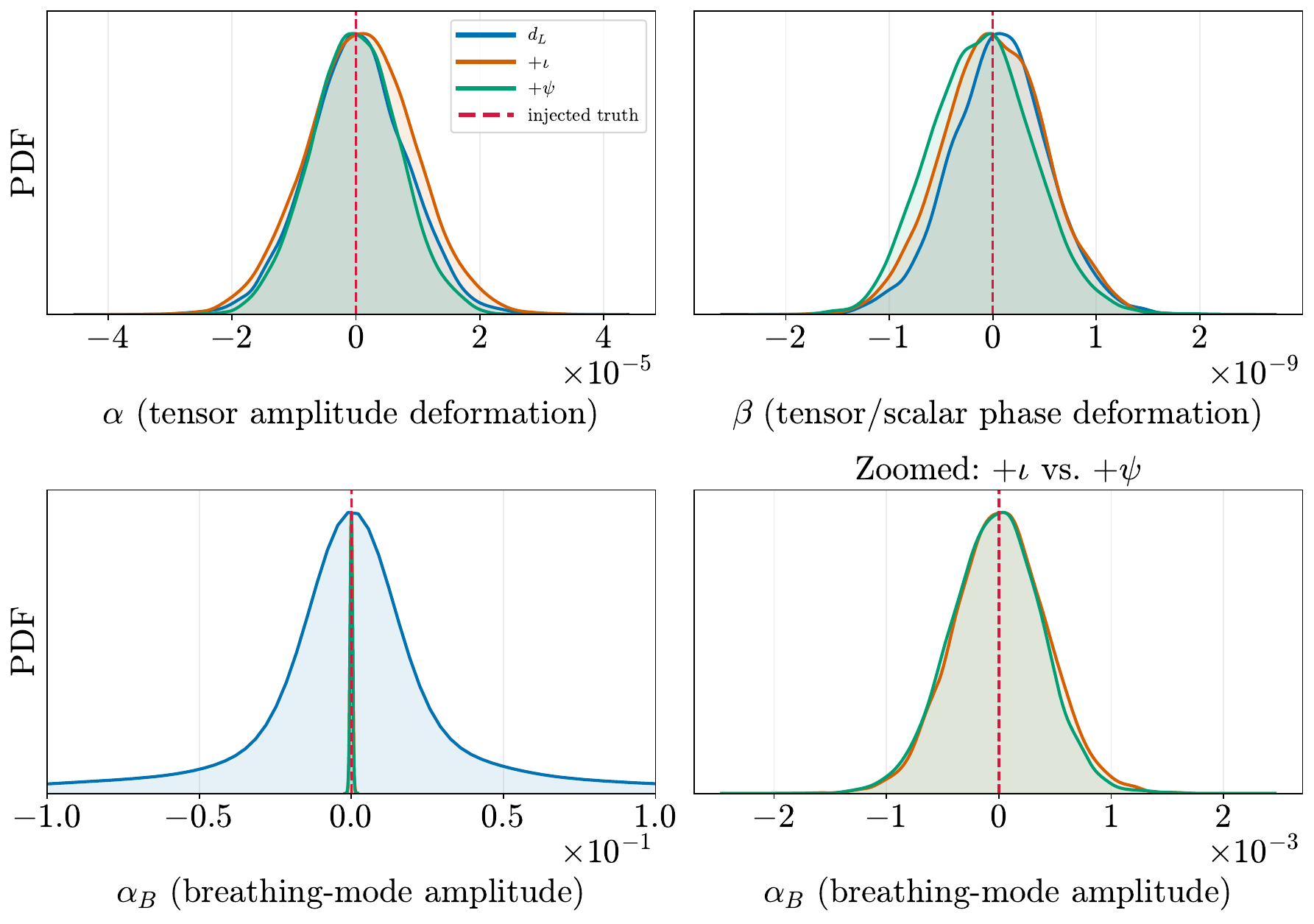}
    \caption{Posterior comparison of $\alpha,\bT,\aB$ across the three
cumulative-prior stages, 2$\times$CE network. $\alpha$ and $\bT$ are
essentially unchanged across all three stages. On $\aB$, adding an
informative $\iota$ prior (Stage 1$\to$2) collapses the 68\% CI by
$\approx56\times$; adding $\psi$ on top (Stage 2$\to$3) barely moves it further, $\approx 1\times$, for this $\alpha_B=0$ injection.  }
    \label{fig:result2}
\end{figure*}

In this section, we reanalyze the GW170817-like mock event, varying the available EM information. In particular, we will always assume the sky location and the luminosity distance $d_L$ to be known from EM observations (obtained from the observation of a kilonova). However, since a collimated jet is more difficult to observe and requires preferential inclination, we consider three scenarios: i) there is no observation from a jet and hence no EM information on $(\iota,\psi)$; (ii) there is initial observation of a jet, which provides information on $\iota$; (iii) there is observation and long-term follow-up of a jet, which provides information on both $(\iota,\psi)$. We perform this analysis only for the $2\times$CE detector network.

Fig.~\ref{fig:result2} shows the posterior probability distributions for the three PPE parameters, in the three scenarios mentioned above, while Table~\ref{tab:result3} quotes the constrained values in each case. Regarding the tensor modifications $(\alpha,\beta)$ we find the three cases (i), (ii), and (iii) to yield similar constraints, which indicates that the observation of a jet is not strictly necessary for testing these deviations from GR. However, for the scalar breathing mode, we find that moving from case (i) (sky+$d_L$ EM information included) to case (ii) ($+\iota$) improves
the $\aB$ 68\% CI by a factor of $\approx56$, from $2.3\times10^{-2}$ to
$4.0\times10^{-4}$, while the inclusion of $\psi$ in case (iii) maintains the constraining precision on $\alpha_B$.
This isolates the addition of inclination information --- on top of
the sky- and $d_L$-informed baseline used throughout this section ---
as the parameter responsible for the $\alpha_B$ sensitivity gain,
consistent with the known degeneracy between the breathing-mode
amplitude and the inclination angle.

\begin{figure}[h!]
     \centering
     \includegraphics[width=0.45\textwidth]{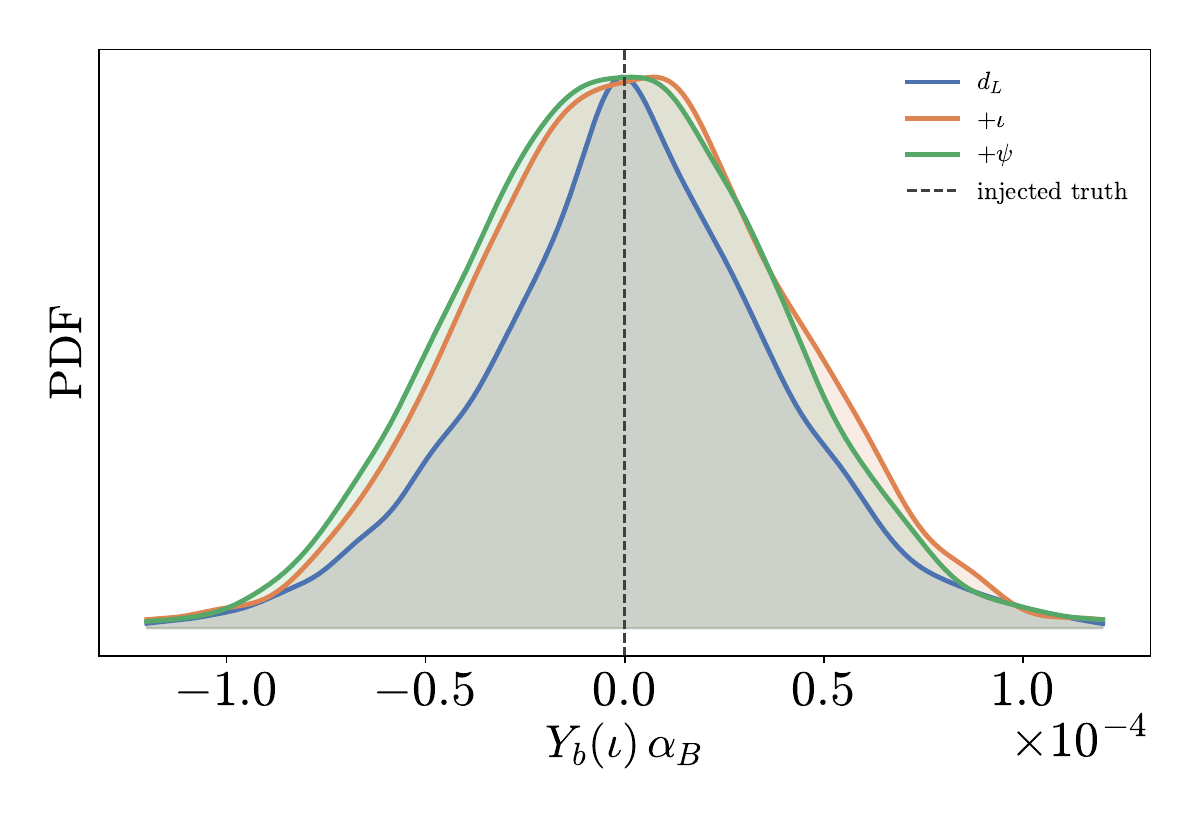}
     \caption{Distribution of $Y_b(\iota)\,\aB$ --- the combination the
breathing-mode strain actually constrains, $h_b\propto Y_b(\iota)\,\aB$ ---
across the three stages of Fig.~\ref{fig:result2}. Its 68\% CI half-width
stays within $\sim3.5$--$3.9\times10^{-5}$ at every stage, unlike $\aB$
alone, which narrows by $\approx57\times$. }
     \label{fig:Yb_alphaB_em}
 \end{figure}

This degeneracy is made explicit in Fig.~\ref{fig:Yb_alphaB_em}, which shows the
distribution of $Y_b(\iota)\,\aB$ --- the combination the GW
data actually constrains --- across all three stages; its width stays
essentially constant ($\sim3.5$--$3.9\times10^{-5}$) even though the raw
$\aB$ posterior narrows by $\approx56\times$ over the same stages. Because $\alpha_B$ is centered on zero for every $\iota$ in our $\alpha_B=0$ injection, $\alpha_B$ and $\iota$ are linearly uncorrelated ($|\rho|\le0.12$ in all stages), yet strongly dependent: $\iota$ sets the width of $\alpha_B$, not its median (for $\alpha_B\neq0$ see Sec.~III\,C). The width satisfies $\sigma(\alpha_B)\sin^2\iota\simeq(3.0$--$3.4)\times10^{-5}$ in all three stages, i.e.\ $\sigma(\alpha_B)\propto1/\sin^2\iota$. With only the sky position and $d_L$ known (Stage~1), $\iota$ is unconstrained and drifts towards the pole (median $\sin^2\iota=0.002$), so the $\alpha_B$ posterior is wide. Once the $\iota$ prior is added (Stages~2 and~3), $\sin^2\iota=0.079\pm0.011$ is nearly fixed, $Y_b\simeq0.10$, and $\sigma(\alpha_B)\simeq\sigma(Y_b\alpha_B)/Y_b\approx4\times10^{-4}$. The GW data constrain only $d_L/f(\iota)$, with $f$ nearly flat near the pole, so $\iota$ and $d_L$ are strongly correlated ($\rho\simeq0.95$--$0.99$) and $\iota$ is not constrained independently of $d_L$ by GW data alone.

\begin{table*}[t]
\begin{ruledtabular}
\caption{Recovered PPE posteriors under three literature-motivated $\iota$-prior
scenarios, 2$\times$CE network. \texttt{HK} is the pipeline's own fiducial
$\iota$-prior run, Gaussian$(2.8495,0.03)$ --- the same informative prior used
as the default throughout this work (Table~\ref{tab:priors}); its internal label follows this prior's origin in the
VLBI superluminal-motion-based inclination measurement of GW170817~\cite{Hotokezaka2019H0}), which motivated the fiducial
injection's $\iota$ value used throughout this work. Ghirlanda~\cite{Ghirlanda2019StructuredJet}
and Wang~\cite{Wang2023ViewingAngle} impose the respective papers' inclination constraint as the assumed $\iota$
prior instead.}
\begin{tabular}{lccc}
Scenario & $\alpha$ & $\bT$ & $\aB$ \\
\hline
HK & $(-0.15\pm6.94)\times10^{-6}$ & $(-1.01\pm4.90)\times10^{-10}$ & $(-0.18\pm3.95)\times10^{-4}$ \\
Ghirlanda & $(-0.08\pm7.17)\times10^{-6}$ & $(0.10\pm4.77)\times10^{-10}$ & $(-0.87\pm5.19)\times10^{-4}$ \\
Wang & $(0.01\pm6.74)\times10^{-6}$ & $(-0.49\pm4.79)\times10^{-10}$ & $(1.63\pm2.34)\times10^{-4}$ \\
\end{tabular}
\end{ruledtabular}
\label{tab:result4}
\end{table*}

\begin{figure*}
    \centering
    \includegraphics[width=0.9\textwidth]{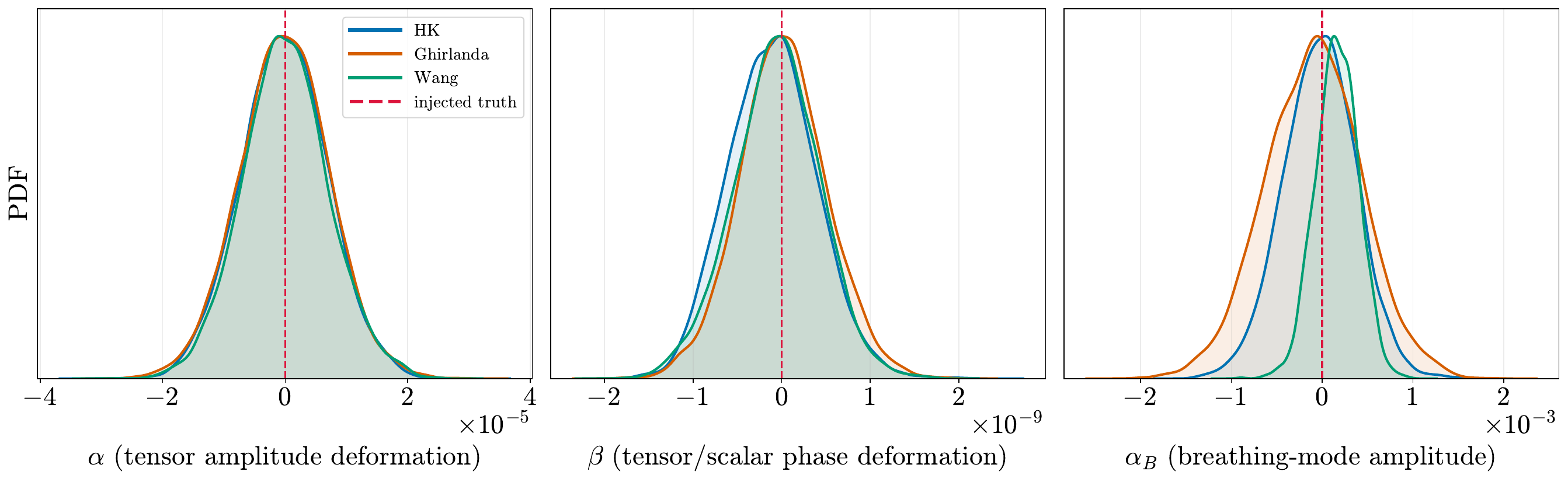}
    \caption{Posterior comparison of $\alpha,\bT,\aB$ across the HK,
Ghirlanda, and Wang $\iota$-prior scenarios (2$\times$CE network). All three
are comparable width on $\alpha$/$\bT$; on $\aB$, Ghirlanda's prior widens the CI ($D_{JS}$(HK,Ghirlanda)=0.029), whereas Wang's tightens it and shifts the median furthest from zero ($D_{JS}$(HK,Wang)=0.123). }
    \label{fig:result3}
\end{figure*}
 
Adding the $\psi$ prior on top barely moves $\aB$ further. This is specific to our $\aB=0$ injection, which sits at the GR value. The fixed point of the $(\psi,\aB)\to(\psi+\pi/2,-\aB)$ sign-flip degeneracy of Ref.~\cite{Imam_2026} becomes irrelevant: the two mirror solutions share the same $\aB$, so no bimodality appears in the $\aB$ posterior at any stage in this injection, and a $\psi$ prior has little to remove. For $\aB\neq0$ the mirror solution is distinct, and a prior on $\psi$ can break this degeneracy, as found for the real GW170817 data in Ref.~\cite{Imam_2026}; quantifying this effect for forecasts is left to future work.
Taken together, the network-driven precision gains of Sec.~\ref{sec:result1} and the prior-driven gains identified here are largely independent, and for $\aB$ specifically the EM-driven gain is dominated by whether an informative $\iota$ prior is available at all; for this $\aB=0$ injection, $\psi$ contributes comparatively little.


\subsection{GW170817: EM Inclination Bias on Breathing Polarization}
\label{Sec:result3}

In this section we test whether the recovered constraint on the breathing-mode
amplitude $\aB$ depends on which astrophysically-inferred inclination is used
as the $\iota$ prior, holding the injection, network ($2\times$CE), and the
sky-, $d_L$-, and $\psi$-informed baseline of Sec.~\ref{sec:result2} fixed.
GW170817's inclination $\iota$ (equivalently the jet viewing angle $\theta_v$,
with $\iota=180^\circ-\theta_v$) has been inferred independently by several
groups under different jet-modeling assumptions, with disagreeing results: the
VLBI superluminal-motion analysis~\cite{Hotokezaka2019H0} (``HK") gives
$\theta_v\approx0.29$ rad, i.e.\ $\iota\approx163.3^\circ$ --- the pipeline's
fiducial prior, Gaussian$(2.8495,0.03)$, also used to set the fiducial
injection; a structured-jet fit to the VLBI image and light
curve~\cite{Ghirlanda2019StructuredJet} gives
$\theta_v=15^{+1.5}_{-1.0}{}^\circ$, i.e.\ $\iota=165^{+1.0}_{-1.5}{}^\circ$
($2.880$ rad); and a late-time afterglow fit~\cite{Wang2023ViewingAngle} gives
$\theta_v=0.53\pm0.01$ rad, i.e.\ $\iota\approx149.6^\circ$ ($2.612$ rad).
These priors differ by up to $\approx16^\circ$ in $\iota$; since $\alpha_B$ enters only through the combination $Y_b(\iota)\,\alpha_B$, we repeat the analysis with each of the HK, Ghirlanda, and
Wang priors in turn to test robustness.

Table~\ref{tab:result4} and Fig.~\ref{fig:result3} isolate the effect of
the assumed $\iota$ prior alone, holding the network (2$\times$CE) and
injection fixed. $\alpha$ and $\bT$ are essentially unaffected by which
$\iota$ prior is assumed --- their 68\% CIs vary by less than 10\% across
all three scenarios ($6.7$--$7.2\times10^{-6}$ for $\alpha$,
$4.8$--$4.9\times10^{-10}$ for $\bT$) --- as expected, since these
tensor-sector deformations enter through GR's own well-constrained
$+/\times$ polarization response and do not share the amplitude-angle
degeneracy discussed below.

$\aB$ tells a different, non-monotonic story. Wang's prior tightens the
$\aB$ 68\% CI relative to the Hotokezaka (``HK") prior --- the pipeline's own fiducial
$\iota$ prior, Gaussian$(2.8495,0.03)$, used as the default throughout this
work (Table~\ref{tab:priors}) --- ($\pm2.3\times10^{-4}$
vs.\ $\pm3.9\times10^{-4}$, $D_{JS}$(HK,Wang)$=0.123$), but Ghirlanda's actually \emph{widens} it
($\pm5.2\times10^{-4}$, $D_{JS}$(HK,Ghirlanda)$=0.029$) with medians of $-8.7\times10^{-5}$ (Ghirlanda), $-1.8\times10^{-5}$ (HK) and $1.6\times10^{-4}$ (Wang), all within their 68\% CIs of the injected GR value. This
traces directly to the leading-order degeneracy $h_b\propto
Y_b(\iota)\,\aB$, in which only the product is constrained by the data:
the recovered $\iota$ posteriors sit at $163.5^\circ$, $165.3^\circ$, and
$158.5^\circ$ for HK, Ghirlanda, and Wang respectively, all close to
the $\iota=180^\circ$ pole where $Y_b(\iota)=\sqrt{3/2}\sin^2\iota\to0$.
Ghirlanda's prior happens to pull the posterior closest to that pole
($Y_b=0.079$, versus $0.099$ for HK and $0.165$ for Wang), which
amplifies rather than suppresses the degenerate uncertainty on $\aB$
despite Ghirlanda's formally tighter $\iota$ constraint. The point is
that an $\iota$ prior's leverage on $\aB$ depends on where it places the
recovered inclination relative to the pole, not simply on how narrow the
prior itself is.
For an injection with $\alpha_B\neq0$ the posterior median also tracks $\iota$: since the data fix $Y_b\alpha_B$, $\alpha_B\propto1/Y_b(\iota)$, giving a genuine correlation between the two parameters (for the GW170817 best-fit injection, $\alpha_B^{\rm inj}=-0.237$, we find $\rho(\alpha_B,\iota)=-0.997$ and $Y_b\alpha_B=-0.024$ to $0.4\%$). To leading order, adopting the Ghirlanda (Wang) $\iota$ prior in place of HK would shift $\alpha_B$ from $-0.23$ to $\approx-0.31$ ($\approx-0.15$).

\subsection{GW170817: Equation-of-State Sensitivity of the PPE Constraints}
\label{sec:eos-sensitivity}

\begin{table*}[t]
\begin{ruledtabular}
\caption{Recovered PPE posteriors (median $\pm$ 68\% credible-interval half-width) for the GW170817-like BNS injection (2$\times$CE network) under three choices of injected tidal deformability, all other settings held unchanged.}
\label{tab:eos-sensitivity}
\begin{tabular}{lccc}
Injected $(\Lambda_1,\Lambda_2)$ & $\alpha$ & $\beta$ & $\alpha_B$ \\
\hline
CE baseline, $(757,1019)$                                          & $(-0.15\pm 6.94)\times10^{-6}$ & $(-1.01\pm 4.90)\times10^{-10}$ & $(-0.18\pm 3.95)\times10^{-4}$ \\
MS1b (stiff), $(369.4,4175.6)$  & $(0.18\pm 6.97)\times10^{-6}$  & $(1.18\pm 4.64)\times10^{-10}$  & $(-0.03\pm 4.07)\times10^{-4}$ \\
SLy4(soft), $(62.7,1353.1)$              & $(0.26\pm 6.98)\times10^{-6}$  & $(5.58\pm 5.03)\times10^{-10}$  & $(-0.18\pm 3.99)\times10^{-4}$ \\
\end{tabular}
\end{ruledtabular}
\end{table*}

\begin{figure*}[t]
\centering
\includegraphics[width=\textwidth]{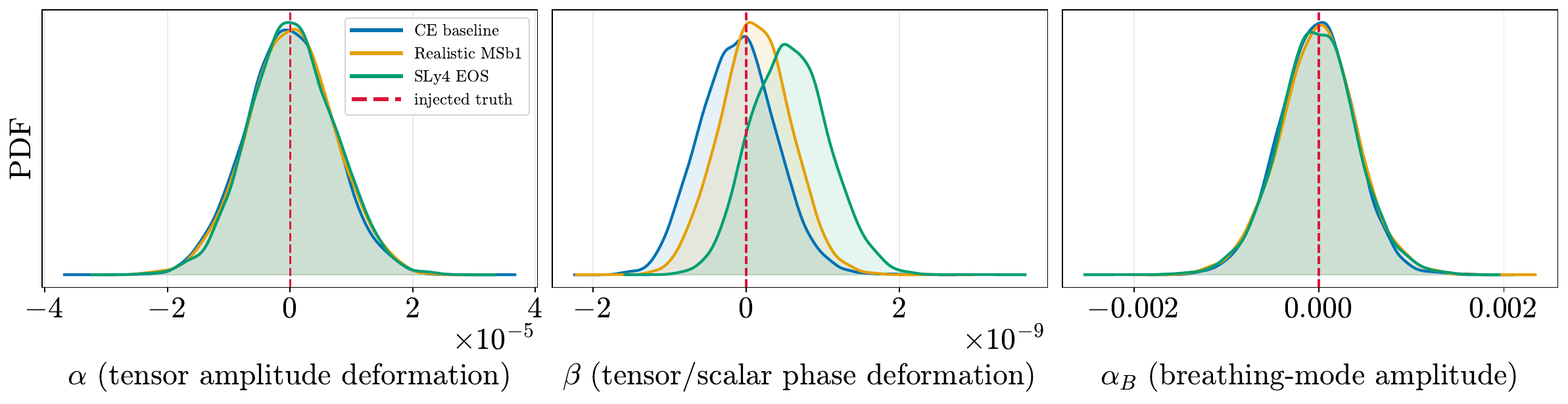}
\caption{Posterior comparison of $\alpha,\beta,\alpha_B$ for the GW170817-like BNS injection under three choices of injected tidal deformability — an initial placeholder (CE baseline), the MS1b EOS (stiff), and the SLy4 EOS (soft) — 2$\times$CE network. Curves are peak-normalized. $\alpha$ and $\alpha_B$ are essentially indistinguishable across all three; $\beta$ shows a mild, sub-dominant shift traceable to its correlation with the recovered $\tilde\Lambda$ (see text).}
\label{fig:eos-sensitivity}
\end{figure*}

In this section we test whether the PPE constraints reported above are
sensitive to the true nuclear equation of state (EOS), by repeating the
analysis with three different injected tidal-deformability pairs
$(\Lambda_1,\Lambda_2)$ --- spanning a stiff and a soft EOS --- while
holding fixed the same wide, EOS-agnostic prior on $\Lambda_1,\Lambda_2$
used throughout this work.

Throughout Secs.~\ref{sec:result1}--\ref{Sec:result3}, the tidal deformabilities $\Lambda_1,\Lambda_2$ are treated as free nuisance parameters with a wide, EOS-agnostic prior (Table~\ref{tab:priors}) that is identical across every analysis in this work and never informed by any specific nuclear equation of state (EOS). The PPE deformation parameters modify the same waveform: $\alpha$ and $\alpha_B$ deform the tensor- and breathing-mode \emph{amplitudes}, respectively, while $\beta$ deforms the tensor-sector \emph{phase}, each through its own fixed power of the frequency-dependent factor $u_m$ (Sec.~\ref{sec:waveform}). Since $\Lambda_1,\Lambda_2$ also enter the waveform through a frequency-dependent term, a natural concern is whether marginalizing over $\Lambda_1,\Lambda_2$ with this fixed, generic prior is adequate regardless of which physically realizable EOS the \emph{true} (injected) tidal deformabilities happen to correspond to. We test this directly: the prior on $\Lambda_1,\Lambda_2$ is held fixed at Table~II's Uniform$(0,5000)$ prior in every run below, while only the \emph{injected} tidal-deformability pair is changed.

We compare three otherwise identical injections of the same  GW170817-like ($\mathcal{M}_c = 1.1975174\,M_\odot$, $\eta = 0.2379$; component masses $m_1 = 1.7289\,M_\odot$, $m_2 = 1.1054\,M_\odot$), analyzed with the 2$\times$CE network, $n_{\rm live}=2000$, with a GR-consistent ($\alpha=\beta=\alpha_B=0$) injection as elsewhere in this work. The three runs differ only in the injected tidal-deformability pair: an initial placeholder choice, $(\Lambda_1,\Lambda_2)=(757,1019)$ (``CE baseline''); the MS1b EOS~\cite{MullerSerot1996,ReadEtAl2009}, a stiff EOS, $(369.4,4175.6)$; and the SLy4 EOS~\cite{DouchinHaensel2001}, a soft EOS, $(62.7,1353.1)$.

The SLy4 (soft) and MS1b (stiff) tidal-deformability pairs are both obtained by solving the Tolman--Oppenheimer--Volkoff~\cite{Tolman1939,OppenheimerVolkoff1939} equations together with the quadrupolar tidal Love-number equation~\cite{Hinderer2010}: the SLy4 unified nuclear-matter EOS~\cite{DouchinHaensel2001} for the soft pair, and the standard piecewise-polytrope fit~\cite{ReadEtAl2009} of the MS1b relativistic mean-field EOS~\cite{MullerSerot1996} for the stiff pair. At the actual injected component masses, this calculation reproduces the pipeline's injected MS1b pair, $(\Lambda_1,\Lambda_2)=(369.4,4175.6)$, exactly, and gives $(\Lambda_1,\Lambda_2)=(62.7,1353.1)$ for SLy4.

Table~\ref{tab:eos-sensitivity} and Fig.~\ref{fig:eos-sensitivity} compare the recovered posteriors. The tensor-amplitude and breathing-mode parameters are essentially unchanged across all three tidal-deformability choices: the 68\% CI half-width on $\alpha$ varies by $\lesssim 0.6\%$ (from $\pm 6.94\times10^{-6}$ to $\pm 6.98\times10^{-6}$) and on $\alpha_B$ by $\lesssim 3\%$ (from $\pm 3.95\times10^{-4}$ to $\pm 4.07\times10^{-4}$), with correspondingly small pairwise Jensen-Shannon divergences, $D_{JS}\lesssim 2\times10^{-3}$ for both parameters across all three pairs (Table~\ref{tab:eos-djs}); the posterior correlation of $\alpha$ and $\alpha_B$ with the leading-order effective tidal deformability $\tilde\Lambda$~\cite{Favata2014,WadeEtAl2014} is negligible ($|r|\lesssim0.05$) in every run, consistent with $\alpha,\alpha_B$ entering the waveform through the amplitude rather than the phase. The tensor/scalar phase parameter $\beta$ shows a somewhat larger, but still sub-dominant, sensitivity: its 68\% CI half-width varies by $\approx 8\%$ (from $\pm 4.64\times10^{-10}$ to $\pm 5.03\times10^{-10}$) and its pairwise $D_{JS}$ ranges from $0.040$ to $0.270$. This is traceable to a real, if sub-dominant, correlation channel: $\beta$ is consistently correlated with the recovered $\tilde\Lambda$ in every run ($r=0.52$--$0.63$), since both enter the frequency-domain phase, whereas the individual $\Lambda_1,\Lambda_2$ posteriors are comparatively prior-dominated and correlate with $\beta$ only weakly and with opposite sign ($r=0.09$--$0.27$ and $-0.07$ to $-0.22$, respectively) — consistent with the well-known result that $\tilde\Lambda$, not $\Lambda_1,\Lambda_2$ individually, is the tidal combination that is robustly measured~\cite{Favata2014,WadeEtAl2014}. In all three runs and for all three parameters, the recovered posteriors remain consistent with the injected GR value $\alpha=\beta=\alpha_B=0$.

These results indicate that the PPE constraints reported throughout this work are not materially degraded by the assumed nuclear equation of state, despite a real but small phase-channel correlation between $\beta$ and the tidal/mass-sector parameters: recovering $\alpha,\beta,\alpha_B$ with the same fixed, EOS-agnostic prior on $\Lambda_1,\Lambda_2$ (Table~II) gives bounds that remain consistent with GR and with each other across a stiff and a soft EOS, supporting the robustness of the results in Secs.~\ref{sec:result1}--\ref{Sec:result3} against the choice of EOS.

\begin{table}[h!]
\begin{ruledtabular}
\caption{Pairwise Jensen-Shannon divergence $D_{JS}$ (bits) between the three tidal-deformability choices of Table~\ref{tab:eos-sensitivity}.}
\label{tab:eos-djs}
\begin{tabular}{lccc}
Parameter & CE base.\,--\,MS1b & CE base.\,--\,SLy4 & MS1b\,--\,SLy4 \\
\hline
$\alpha$    & $9.0\times10^{-4}$ & $2.0\times10^{-3}$ & $9.2\times10^{-4}$ \\
$\beta$     & $0.040$            & $0.270$            & $0.141$            \\
$\alpha_B$  & $1.9\times10^{-3}$ & $2.0\times10^{-3}$ & $9.2\times10^{-4}$ \\
\end{tabular}
\end{ruledtabular}
\end{table}

It is worth emphasizing that this robustness is demonstrated only within
the inspiral-only, quadrupole-only PPE analysis adopted throughout this
work, restricted to $f\le f_c^{(2,2)}$ with $M_zf_c^{(2,2)}=0.018$
(Sec.~\ref{sec:waveform}); tidal effects on the phase grow rapidly with
frequency and are strongest in the final orbits and merger, precisely the
band this cutoff excludes~\cite{Hinderer2010,Favata2014}. Full
inspiral-merger-ringdown tests of GR of the kind routinely performed by
the LIGO--Virgo--KAGRA collaboration~\cite{LIGOScientific:2021sio} (see
also the TIGER framework~\cite{Agathos:2013upa}) instead include the merger
band, where the tidal-EOS coupling is strongest and the compact objects'
internal structure directly shapes the waveform; we therefore cannot
conclude from the present, inspiral-restricted result that $\aB$ (or the
tensor-sector $\alpha,\bT$) would remain similarly insensitive to the EOS
in a full-IMR analysis, and we flag this as a limitation of the
inspiral-only PPE framework adopted here rather than a general statement
about scalar-polarization tests of GR.

\subsection{Cross-Detector Comparison for a Representative BNS Event}
\label{sec:result4}

In this section we repeat the cross-network comparison of
Sec.~\ref{sec:result1} for a second, independent BNS injection ---
taken from the future representative-event assumptions of
\cite{Lagos:2024boe} --- to test whether the network-precision ordering
found for the fiducial GW170817-like injection generalizes to a more
typical detection. GW170817 was itself an outlier: its proximity
($d_L\approx42\,$Mpc) and near-face-on orientation ($\theta_v\approx17^\circ$,
i.e.\ $\iota\approx2.8495\,$rad) are atypical of the BNS population
next-generation detectors are expected to catalog, since detection volume
grows with distance while only a small solid-angle fraction of random
orientations lies this close to face-on. This second injection instead
places the source at a more representative distance ($d_L=100\,$Mpc) and
a larger viewing angle ($\theta_v=30^\circ$, i.e.\ $\iota=2.62\,$rad) ---
at the cost of a lower network SNR and a correspondingly fainter,
more off-axis GRB counterpart, the same distance/inclination-vs.detectability
trade-off that already made GRB\,170817A itself an unusually faint
short GRB even at its comparatively face-on viewing angle.

\begin{figure*}
    \centering
    \includegraphics[width=0.9\textwidth]{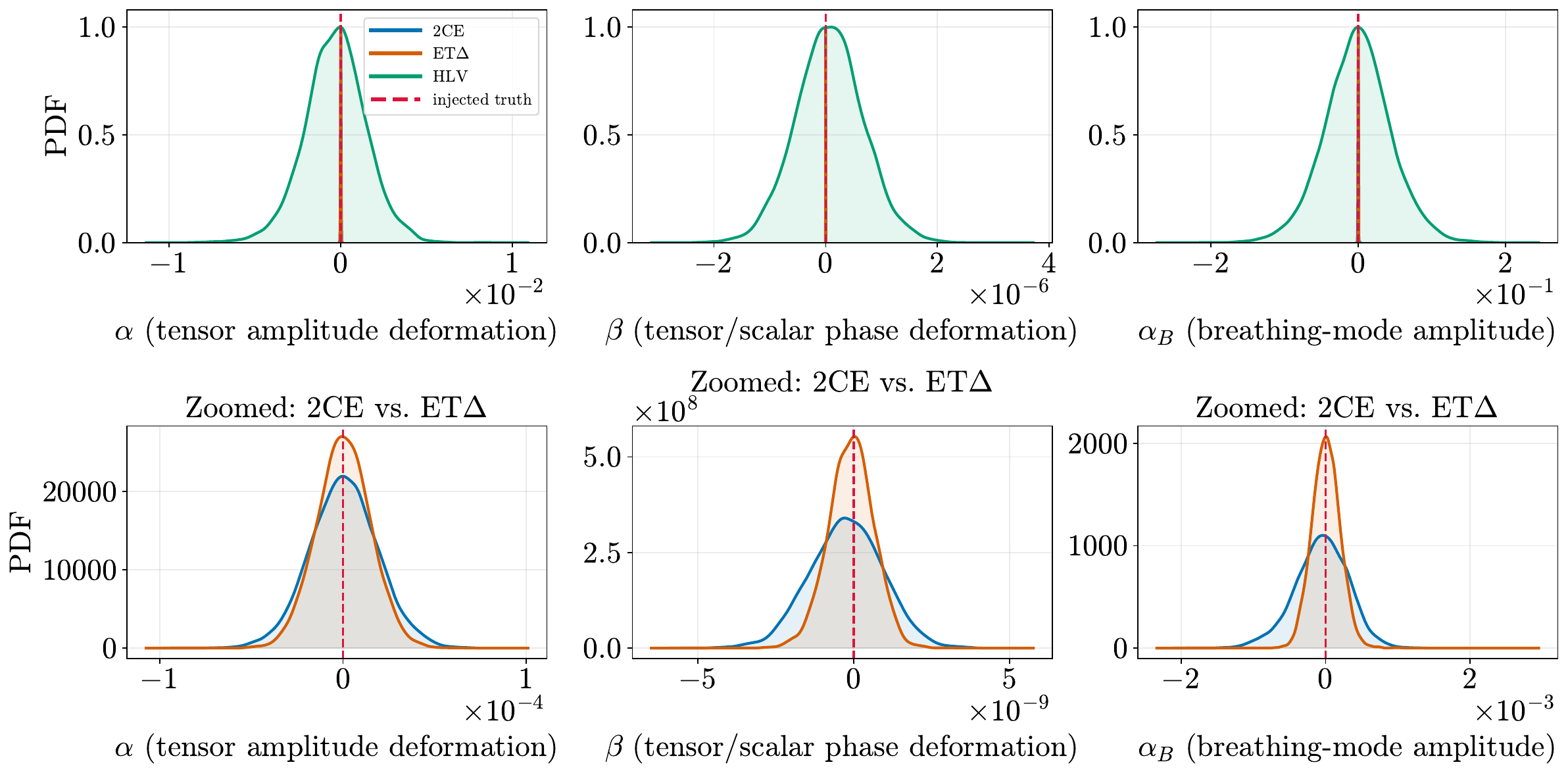}
    \caption{BNS representative event:  Distribution of PPE parameters across different detectors.}
    \label{fig:result4}
\end{figure*}

Table~\ref{tab:result5} and Fig.~\ref{fig:result4} repeat the network
comparison of Sec.~\ref{sec:result1} for a second, independent
representative BNS injection, testing whether the precision ordering found
there generalizes. As before, O5/HLV is dramatically outperformed by both
next-generation networks on every parameter: its 68\% CI is
$\approx93\times$ wider than 2$\times$CE's on $\alpha$
($\pm1.7\times10^{-3}$ vs.\ $\pm1.8\times10^{-5}$) and $\approx225\times$
wider than ET's on $\aB$ ($\pm4.3\times10^{-2}$ vs.\ $\pm1.9\times10^{-4}$),
consistent with the analogous network-sensitivity gap in Sec.\ref{sec:result1}.

The 2$\times$CE-vs-ET$\Delta$ ordering, however, does \emph{not} generalize. In
Sec.~\ref{sec:result1}, CE was tighter than ET$\Delta$ on $\alpha$ and $\bT$, with ET$\Delta$
only pulling ahead $\approx2\times$ on $\aB$. Here, ET$\Delta$ is tighter than
2$\times$CE on all three parameters simultaneously: $\approx1.3\times$ on
$\alpha$ ($\pm1.4\times10^{-5}$ vs.\ $\pm1.8\times10^{-5}$
), $\approx1.6\times$ on $\bT$
($\pm7.0\times10^{-10}$ vs.\ $\pm1.2\times10^{-9}$
), and $\approx1.9\times$ on $\aB$
($\pm1.9\times10^{-4}$ vs.\ $\pm3.6\times10^{-4}$
). This event-to-event
reversal on $\alpha$/$\bT$ indicates that CE and ET's relative ranking on the
tensor-sector parameters is not fixed by network geometry alone --- it also
depends on the specific source configuration (sky location, inclination,
and the resulting relative SNR  $\approx 471$ for ET$\Delta$ and  $\approx 1184$ for 2$\times$CE split between the two networks for this
particular injection) --- while ET's advantage on $\aB$ persists across
both events, consistent with its triangular geometry's polarization-mode
discrimination being the more robust driver for that parameter
specifically.

\begin{table*}[t]
\begin{ruledtabular}
\caption{Representative-BNS-event posteriors: median and 68\% CI half-width
on each PPE parameter, by network.}
\label{tab:result5}
\begin{tabular}{lccc}
Network & $\alpha$ & $\bT$ & $\aB$ \\
\hline
O5/LVK & $(-0.28\pm1.68)\times10^{-3}$ & $(0.72\pm5.88)\times10^{-7}$ & $(-0.06\pm4.25)\times10^{-2}$ \\
2$\times$CE & $(0.04\pm1.81)\times10^{-5}$ & $(-0.23\pm1.16)\times10^{-9}$ & $(-0.43\pm3.57)\times10^{-4}$ \\
ET$\Delta$ & $(0.04\pm1.44)\times10^{-5}$ & $(-0.68\pm7.05)\times10^{-10}$ & $(0.07\pm1.89)\times10^{-4}$ \\
\end{tabular}
\end{ruledtabular}
\end{table*}

\section{ Conclusion}
\label{sec:conclusion}

In this paper, we forecast constraints on GW scalar polarizations, using multi-messenger binary neutron star events. We discuss different detector sensitivities, different EM informative priors, and assess possible biases due to mismodeling of the EM jet or neutron star equation of state.

The results found trace a single thread: how precisely a
GR-consistent binary neutron star merger constrains the
post-Einsteinian parameters $\alpha$, $\bT$, and $\aB$ depends not only
on the detector network, but --- for $\aB$ specifically --- on how much
external information the analysis assumes about the source's orbital
inclination. A network's raw sensitivity (Sec.~\ref{sec:result1}) sets
one axis of precision: both next-generation networks outperform O5/LVK
by 1--2 orders of magnitude on every PPE parameter, and CE and ET$\Delta$ are
not interchangeable points on a single SNR axis --- CE's higher
per-detector strain sensitivity gives it the edge on the tensor-sector
parameters $\alpha$ and $\bT$, while ET's triangular geometry gives it a
persistent, roughly $2\times$ advantage on $\aB$ specifically, a
distinction that survived the switch to an independent representative
event in Sec.~\ref{sec:result4} even as the $\alpha$/$\bT$ ordering
between CE and ET$\Delta$ reversed.

The second axis, separate from network sensitivity, is prior informativeness on the inclination angle $\iota$. Sec.~\ref{sec:result2}
isolated $\iota$ as the parameter responsible, for this injection, for essentially all of $\alpha_B$'s sensitivity to external priors: adding an informative $\iota$ prior collapsed the $\alpha_B$ credible interval by $\approx 56\times$, while adding a $\psi$ prior on top moved it by only $\approx 1\times$ (for our $\alpha_B=0$ injection, where the $\psi$-mirror degeneracy is trivial; $d_L$ and sky are assumed known throughout). This is the direct, quantitative signature of the leading-order
degeneracy $h_b\propto Y_b(\iota)\,\aB$, in which only the product is
constrained by GW data. Sec.~\ref{Sec:result3} showed that this
leverage is not simply a matter of how narrow the assumed $\iota$ prior
is: a formally tighter, literature-motivated prior (Ghirlanda) produced
a \emph{wider} $\aB$ constraint than the pipeline's own fiducial $\iota$
prior (Gaussian$(2.8495,0.03)$, itself already an informative,
literature-motivated choice, because it
happened to place the recovered inclination closer
to the $\iota=180^\circ$ pole where $Y_b(\iota)\to0$ and the degeneracy
is most severe. An external follow-up campaign that constrains $\iota$
tightly but off to one side of the true value can therefore do more
harm than good for a scalar-polarization bound, a failure mode that a
naive ``narrower prior is always better'' heuristic would miss entirely.

Taken together, these results indicate that next-generation
scalar-polarization constraints will be shaped as much by which detector
geometry and which external inclination information happen to be
available for a given event as by the raw sensitivity gain from O5 to
next-generation networks. A triangular network's polarization-mode
diversity and a well-placed (not merely narrow) inclination constraint
are each, independently, worth as much to an $\aB$ bound as an
order-of-magnitude increase in network SNR --- a consideration that
argues for treating detector-network design and multi-messenger
follow-up strategy as jointly relevant to the scientific case for
next-generation gravitational-wave polarization tests, rather than
optimizing SNR alone.

Beyond these two axes, Sec.~\ref{sec:eos-sensitivity} shows the GW170817-like  results above are not an artifact of the assumed nuclear equation of state: a GW170817-like injection with three different injected tidal deformabilities (a placeholder, the stiff MS1b EOS, and the soft SLy4 EOS) leaves $\alpha$ and $\alpha_B$ essentially unchanged and shifts $\beta$ only mildly, an effect traceable to $\beta$'s correlation with the recovered effective tidal deformability $\tilde\Lambda$ and still more than an order of magnitude smaller than the network- and prior-driven effects identified above   (within the inspiral-only analysis, and not retested for the representative event of Sec.~\ref{sec:result4}).

For context, Ref.~\cite{Akama:2026igu} forecast analogous PPE
parameters for LISA massive black-hole binaries, using Fisher matrices and GR
injections. For frequency scalings matching ours, $(a_T,b,a_S)=(-2,-7,-2)$,
they find $\sigma(\alpha)\sim10^{-4}$--$10^{-3}$ and
$\sigma(\beta)\sim10^{-7}$--$10^{-5}$~rad for sources at $z=1$--$4$ (with
$M\sim10^{6}\,M_\odot$ for $\beta$). For the scalar amplitude with $a_B=-2$ they
obtain $\sigma\sim10^{-3}$, reaching $\sim10^{-4}$ for the $\ell=2$ breathing
amplitude in the most favorable mass and redshift range. For the 2$\times$CE
network, our constraints on $\alpha$ and $\beta$ are more than an order of
magnitude tighter than these LISA values, while our constraint on $\alpha_B$ is
of comparable size (Table~\ref{tab:result1}). These numbers are, however, not
directly comparable. Ref.~\cite{Akama:2026igu} varies one PPE parameter at a
time and quotes medians over random sky positions and inclinations, whereas we
sample $(\alpha,\beta,\alpha_B)$ jointly for a single event with known sky
position and $d_L$ and an informative inclination prior. The breathing
amplitude in particular is suppressed by $\sin^2\iota$, which is small for the
inclination of this event ($\sin^2\iota\simeq0.08$) but is averaged over in the
LISA medians. 
Ref.~\cite{Wu:2024yno} forecast PPE constraints on the vector and
scalar amplitudes with a Fisher matrix for LISA, Taiji, TianQin, LVK and ET.
Ground-based detectors cannot separate the breathing and longitudinal modes, so
their scalar-amplitude errors are four to six orders of magnitude larger than
those of space-based networks. We include only the breathing mode, so our
$\alpha_B$ constraints assume a vanishing longitudinal amplitude. Knowing the
distance, inclination or sky position improves the LISA/Taiji scalar and vector
errors by only $15$--$49\%$, and the scalar errors on the ground not at all
because of that degeneracy, whereas our inclination prior sharpens the
$\alpha_B$ posterior by $\approx56$ (Table~\ref{tab:result3}).

Two limitations bound how far the conclusions of this paper generalize. First, both
the network comparison and the prior study are built on one or two
fiducial injections rather than a source population; Sec.~\ref{sec:result4}'s
event-to-event reversal of the CE/ET$\Delta$ ordering on $\alpha$ and $\bT$ is
itself evidence that single-event conclusions about \emph{which}
next-generation network is preferable should be treated as
illustrative, not definitive, until confirmed across a broader set of
sky positions, inclinations, and distances. Second, we performed only GR injections ($\alpha=\beta=\alpha_B=0$); injections with non-zero modified-gravity parameters may exhibit stronger parameter correlations (e.g.\ between $\alpha_B$ and $\iota$) and multimodal posteriors. Non-GR injections with theory-motivated parameter values could be considered in the future. 

\begin{acknowledgments}
The author thanks Thibeau Wouters and Thomas
C.\ K.\ Ng for helpful discussions on implementing the
BlackJAX-accelerated nested-sampling annealed-walkers
(NS-AW) within the Jim framework. The authors
acknowledges the use of the NLHPC (Centro de
Modelamiento Matemático, Chile) for computational
resources. The work of M.L.\ was partially supported by Fondecyt Iniciación grant No.\ 11250105. 
\end{acknowledgments}
\section{Data Availability}
The code and results supporting the findings of this study publicly available \cite{ImamLagosXGCode2026}. 
The posterior samples of this work are publicly available on Zenodo~\cite{ImamLagosXGZenodo2026}.

\bibliography{refs}

@ARTICLE{LIN_1991,
  author={Lin, J.},
  journal={IEEE Transactions on Information Theory}, 
  title={Divergence measures based on the Shannon entropy}, 
  year={1991},
  volume={37},
  number={1},
  pages={145-151},
  doi={10.1109/18.61115}}

@article{Lagos:2024boe,
    author = "Lagos, Macarena and Jenks, Leah and Isi, Maximiliano and Hotokezaka, Kenta and Metzger, Brian D. and Burns, Eric and Farr, Will M. and Perkins, Scott and Wong, Kaze W. K. and Yunes, Nicolas",
    title = "{Birefringence tests of gravity with multimessenger binaries}",
    eprint = "2402.05316",
    archivePrefix = "arXiv",
    primaryClass = "gr-qc",
    doi = "10.1103/PhysRevD.109.124003",
    journal = "Phys. Rev. D",
    volume = "109",
    number = "12",
    pages = "124003",
    year = "2024"
}

@article{Planck2018,
    author = "Aghanim, N. and others",
    collaboration = "Planck",
    title = "{Planck 2018 results. VI. Cosmological parameters}",
    eprint = "1807.06209",
    archivePrefix = "arXiv",
    primaryClass = "astro-ph.CO",
    doi = "10.1051/0004-6361/201833910",
    journal = "Astron. Astrophys.",
    volume = "641",
    pages = "A6",
    year = "2020",
    note = "[Erratum: Astron.Astrophys. 652, C4 (2021)]"
}

@article{Imam_2026,
  title = {Tests of scalar polarizations with multimessenger events},
  author = {Imam, Sk Md Adil and Lagos, Macarena},
  journal = {Phys. Rev. D},
  pages = {},
  year = {2026},
  month = {Aug},
  publisher = {American Physical Society},
  doi = {10.1103/f9ll-y1rn},
  url = {https://link.aps.org/doi/10.1103/f9ll-y1rn}
}

@article{IacovelliGWFAST2022,
  title = {{GWFAST}: A {F}isher Information Matrix Python Code for Third-Generation Gravitational-Wave Detectors},
  author = {Iacovelli, Francesco and Mancarella, Michele and Foffa, Stefano and Maggiore, Michele},
  journal = {Astrophys. J. Suppl.},
  volume = {263},
  pages = {2},
  year = {2022},
  doi = {10.3847/1538-4365/ac9129},
  eprint = {2207.06910},
  archivePrefix = {arXiv}
}

@article{Srivastava2022CosmicExplorerDesign,
  title = {Science-Driven Tunable Design of Cosmic Explorer Detectors},
  author = {Srivastava, Varun and Davis, Derek and Kuns, Kevin and Landry, Philippe and Ballmer, Stefan and Evans, Matthew and Hall, Evan D. and Read, Jocelyn and Sathyaprakash, B. S.},
  journal = {Astrophys. J.},
  volume = {931},
  pages = {22},
  year = {2022},
  doi = {10.3847/1538-4357/ac5f04},
  eprint = {2201.10668},
  archivePrefix = {arXiv}
}

@article{Branchesi2023ETDesigns,
doi = {10.1088/1475-7516/2023/07/068},
year = {2023},
month = {jul},
publisher = {IOP Publishing},
volume = {2023},
number = {07},
pages = {068},
author = {Branchesi, Marica and others},
title = {Science with the Einstein Telescope: a comparison of different designs},
journal = {Journal of Cosmology and Astroparticle Physics},
}

@article{LIGOScientific:2017zic,
    author = "Abbott, B. P. and others",
    collaboration = "LIGO Scientific, Virgo, Fermi-GBM, INTEGRAL",
    title = "{Gravitational Waves and Gamma-rays from a Binary Neutron Star Merger: GW170817 and GRB 170817A}",
    eprint = "1710.05834",
    archivePrefix = "arXiv",
    primaryClass = "astro-ph.HE",
    reportNumber = "LIGO-P1700308",
    doi = "10.3847/2041-8213/aa920c",
    journal = "Astrophys. J. Lett.",
    volume = "848",
    number = "2",
    pages = "L13",
    year = "2017"
}

@article{Pratten:2020fqn,
    author = "Pratten, Geraint and Husa, Sascha and Garcia-Quiros, Cecilio and Colleoni, Marta and Ramos-Buades, Antoni and Estelles, Hector and Jaume, Rafel",
    title = "{Setting the cornerstone for a family of models for gravitational waves from compact binaries: The dominant harmonic for nonprecessing quasicircular black holes}",
    eprint = "2001.11412",
    archivePrefix = "arXiv",
    primaryClass = "gr-qc",
    reportNumber = "LIGO-P2000018",
    doi = "10.1103/PhysRevD.102.064001",
    journal = "Phys. Rev. D",
    volume = "102",
    number = "6",
    pages = "064001",
    year = "2020"
}

@article{Edwards:2023sak,
  author  = {Edwards, Thomas D. P. and Wong, Kaze W. K. and Lam, Kelvin K. H. and Coogan, Adam and Foreman-Mackey, Daniel and Isi, Maximiliano and Zimmerman, Aaron},
  title   = {Differentiable and hardware-accelerated waveforms for gravitational wave data analysis},
  journal = {Phys. Rev. D},
  volume  = {110},
  number  = {6},
  pages   = {064028},
  year    = {2024},
  doi     = {10.1103/PhysRevD.110.064028},
  eprint  = {2302.05329},
  archivePrefix = {arXiv},
  primaryClass = {astro-ph.IM}
}

@article{Wong:2023lgb,
    author = "Wong, Kaze W. K. and Isi, Maximiliano and Edwards, Thomas D. P.",
    title = "{Fast Gravitational-wave Parameter Estimation without Compromises}",
    eprint = "2302.05333",
    archivePrefix = "arXiv",
    primaryClass = "astro-ph.IM",
    doi = "10.3847/1538-4357/acf5cd",
    journal = "Astrophys. J.",
    volume = "958",
    number = "2",
    pages = "129",
    year = "2023"
}

@article{Wouters:2024oxj,
    author = "Wouters, Thibeau and Pang, Peter T. H. and Dietrich, Tim and Van Den Broeck, Chris",
    title = "{Robust parameter estimation within minutes on gravitational wave signals from binary neutron star inspirals}",
    eprint = "2404.11397",
    archivePrefix = "arXiv",
    primaryClass = "astro-ph.IM",
    doi = "10.1103/PhysRevD.110.083033",
    journal = "Phys. Rev. D",
    volume = "110",
    number = "8",
    pages = "083033",
    year = "2024"
}

@article{Hotokezaka2019H0,
    author = "Hotokezaka, Kenta and Nakar, Ehud and Gottlieb, Ore and Nissanke, Samaya and Masuda, Kento and Hallinan, Gregg and Mooley, Kunal P. and Deller, Adam. T.",
    title = "{A Hubble constant measurement from superluminal motion of the jet in GW170817}",
    eprint = "1806.10596",
    archivePrefix = "arXiv",
    primaryClass = "astro-ph.CO",
    doi = "10.1038/s41550-019-0820-1",
    journal = "Nature Astron.",
    volume = "3",
    number = "10",
    pages = "940--944",
    year = "2019"
}

@article{Ghirlanda2019StructuredJet,
    author = "Ghirlanda, G. and others",
    title = "{Compact radio emission indicates a structured jet was produced by a binary neutron star merger}",
    eprint = "1808.00469",
    archivePrefix = "arXiv",
    primaryClass = "astro-ph.HE",
    doi = "10.1126/science.aau8815",
    journal = "Science",
    volume = "363",
    pages = "968",
    year = "2019"
}

@article{Wang2023ViewingAngle,
    author = "Wang, Yi-Ying and Tang, Shao-Peng and Jin, Zhi-Ping and Fan, Yi-Zhong",
    title = "{The Late Afterglow of GW170817/GRB 170817A: A Large Viewing Angle and the Shift of the Hubble Constant to a Value More Consistent with the Local Measurements}",
    eprint = "2208.09121",
    archivePrefix = "arXiv",
    primaryClass = "astro-ph.HE",
    doi = "10.3847/1538-4357/aca96c",
    journal = "Astrophys. J.",
    volume = "943",
    number = "1",
    pages = "13",
    year = "2023"
}

@article{Yunes_2009,
   title={Fundamental theoretical bias in gravitational wave astrophysics and the parametrized post-Einsteinian framework},
   volume={80},
   ISSN={1550-2368},
   url={http://dx.doi.org/10.1103/PhysRevD.80.122003},
   DOI={10.1103/physrevd.80.122003},
   number={12},
   journal={Physical Review D},
   publisher={American Physical Society (APS)},
   author={Yunes, Nicolás and Pretorius, Frans},
   year={2009},
   month=dec }

@article{Chatziioannou:2012rf,
    author = "Chatziioannou, Katerina and Yunes, Nicolas and Cornish, Neil",
    title = "{Model-Independent Test of General Relativity: An Extended post-Einsteinian Framework with Complete Polarization Content}",
    eprint = "1204.2585",
    archivePrefix = "arXiv",
    primaryClass = "gr-qc",
    doi = "10.1103/PhysRevD.86.022004",
    journal = "Phys. Rev. D",
    volume = "86",
    pages = "022004",
    year = "2012",
    note = "[Erratum: Phys.Rev.D 95, 129901 (2017)]"
}

@article{Akama:2026igu,
    author = "Akama, Shingo and others",
    collaboration = "LISA Cosmology Working Group",
    title = "{Testing gravitational wave polarizations with LISA}",
    eprint = "2603.03165",
    archivePrefix = "arXiv",
    primaryClass = "astro-ph.CO",
    doi = "10.1088/1475-7516/2026/09/055",
    journal = "JCAP",
    volume = "09",
    pages = "055",
    year = "2026"
}

@article{Mezzasoma:2026wme,
    author = "Mezzasoma, Simone and Haster, Carl-Johan and Yunes, Nicol{\'a}s",
    title = "{Impact of numerical-relativity waveform calibration on parametrized post-Einsteinian tests}",
    eprint = "2603.15765",
    archivePrefix = "arXiv",
    primaryClass = "gr-qc",
    doi = "10.1103/y6lc-fr77",
    journal = "Phys. Rev. D",
    volume = "113",
    number = "12",
    pages = "124043",
    year = "2026"
}

@article{Abac:2023ujg,
    author = "Abac, Adrian and Dietrich, Tim and Buonanno, Alessandra and Steinhoff, Jan and Ujevic, Maximiliano",
    title = "{New and robust gravitational-waveform model for high-mass-ratio binary neutron star systems with dynamical tidal effects}",
    eprint = "2311.07456",
    archivePrefix = "arXiv",
    primaryClass = "gr-qc",
    doi = "10.1103/PhysRevD.109.024062",
    journal = "Phys. Rev. D",
    volume = "109",
    number = "2",
    pages = "024062",
    year = "2024"
}

@book{Weinberg1972,
  author    = {Weinberg, S.},
  title     = {Gravitation and Cosmology},
  publisher = {Wiley},
  year      = {1972}
}

@book{MTW1973,
  author    = {Misner, C. W. and Thorne, K. S. and Wheeler, J. A.},
  title     = {Gravitation},
  publisher = {W. H. Freeman},
  year      = {1973}
}

@article{Eardley_PRD,
  title = {Gravitational-Wave Observations as a Tool for Testing Relativistic Gravity},
  author = {Eardley, Douglas M. and Lee, David L. and Lightman, Alan P.},
  journal = {Phys. Rev. D},
  volume = {8},
  issue = {10},
  pages = {3308--3321},
  numpages = {0},
  year = {1973},
  month = {Nov},
  publisher = {American Physical Society},
  doi = {10.1103/PhysRevD.8.3308},
  url = {https://link.aps.org/doi/10.1103/PhysRevD.8.3308}
}

@book{Will_1993, place={Cambridge}, title={Theory and Experiment in Gravitational Physics}, publisher={Cambridge University Press}, author={Will, Clifford M.}, year={1993}}

@book{Maggiore2008,
  author    = {Maggiore, M.},
  title     = {Gravitational Waves, Volume 1: Theory and Experiments},
  publisher = {Oxford Univ. Press},
  year      = {2008}
}

@book{Schutz2009,
  author    = {Schutz, B. F.},
  title     = {A First Course in General Relativity},
  publisher = {Cambridge Univ. Press},
  edition   = {2},
  year      = {2009}
}

@article{Will:2014kxa,
	Archiveprefix = {arXiv},
	Author = {Will, Clifford M.},
	Doi = {10.12942/lrr-2014-4},
	Eprint = {1403.7377},
	Journal = {Living Rev. Rel.},
	Pages = {4},
	Primaryclass = {gr-qc},
	Slaccitation = {%%CITATION = ARXIV:1403.7377;%%},
	Title = {{The Confrontation between General Relativity and Experiment}},
	Volume = {17},
	Year = {2014}}

@article{Punturo:2010zz,
    author = "Punturo, M. and others",
    editor = "Ricci, Fulvio",
    title = "{The Einstein Telescope: A third-generation gravitational wave observatory}",
    doi = "10.1088/0264-9381/27/19/194002",
    journal = "Class. Quant. Grav.",
    volume = "27",
    pages = "194002",
    year = "2010"
}

@article{Reitze:2019iox,
    author = "Reitze, David and others",
    title = "{Cosmic Explorer: The U.S. Contribution to Gravitational-Wave Astronomy beyond LIGO}",
    reportNumber = "LIGO-P1900316",
    journal = "Bull. Am. Astron. Soc.",
    volume = "51",
    number = "7",
    pages = "035",
    year = "2019"
}

@article{Bekenstein_2004,
  title = {Relativistic gravitation theory for the modified Newtonian dynamics paradigm},
  author = {Bekenstein, Jacob D.},
  journal = {Phys. Rev. D},
  volume = {70},
  issue = {8},
  pages = {083509},
  numpages = {28},
  year = {2004},
  month = {Oct},
  publisher = {American Physical Society},
  doi = {10.1103/PhysRevD.70.083509},
  url = {https://link.aps.org/doi/10.1103/PhysRevD.70.083509}
}

@INPROCEEDINGS{Bekenstein_2006,
       author = {{Bekenstein}, J.~D. and {Sanders}, R.~H.},
        title = "{A Primer to Relativistic MOND Theory}",
    booktitle = {EAS Publications Series},
         year = 2006,
       series = {EAS Publications Series},
       volume = {20},
        month = jan,
    publisher = {EDP},
        pages = {225-230},
          doi = {10.1051/eas:2006075},
archivePrefix = {arXiv},
       eprint = {astro-ph/0509519},
 primaryClass = {astro-ph},
       adsurl = {https://ui.adsabs.harvard.edu/abs/2006EAS....20..225B}
}

@article{Jacobson_2004,
  title = {Einstein-aether waves},
  author = {Jacobson, T. and Mattingly, D.},
  journal = {Phys. Rev. D},
  volume = {70},
  issue = {2},
  pages = {024003},
  numpages = {5},
  year = {2004},
  month = {Jul},
  publisher = {American Physical Society},
  doi = {10.1103/PhysRevD.70.024003},
  url = {https://link.aps.org/doi/10.1103/PhysRevD.70.024003}
}

@misc{eling2005,
      title={Einstein-Aether Theory}, 
      author={C. Eling and T. Jacobson and D. Mattingly},
      year={2005},
      eprint={gr-qc/0410001},
      archivePrefix={arXiv},
      primaryClass={gr-qc},
      url={https://arxiv.org/abs/gr-qc/0410001}, 
}

@misc{jacobson2008,
      title={Einstein-aether gravity: a status report}, 
      author={Ted Jacobson},
      year={2008},
      eprint={0801.1547},
      archivePrefix={arXiv},
      primaryClass={gr-qc},
      url={https://arxiv.org/abs/0801.1547}, 
}

@article{Lightman_73,
  title = {New Two-Metric Theory of Gravity with Prior Geometry},
  author = {Lightman, Alan P. and Lee, David L.},
  journal = {Phys. Rev. D},
  volume = {8},
  issue = {10},
  pages = {3293--3302},
  numpages = {0},
  year = {1973},
  month = {Nov},
  publisher = {American Physical Society},
  doi = {10.1103/PhysRevD.8.3293},
  url = {https://link.aps.org/doi/10.1103/PhysRevD.8.3293}
}

@article{ROSEN_1974,
title = {A theory of gravitation},
journal = {Annals of Physics},
volume = {84},
number = {1},
pages = {455-473},
year = {1974},
issn = {0003-4916},
doi = {https://doi.org/10.1016/0003-4916(74)90311-X},
url = {https://www.sciencedirect.com/science/article/pii/000349167490311X},
author = {Nathan Rosen},
}

@article{Carson:2019rda,
    author = "Carson, Zack and Yagi, Kent",
    title = "{Multi-band gravitational wave tests of general relativity}",
    eprint = "1905.13155",
    archivePrefix = "arXiv",
    primaryClass = "gr-qc",
    doi = "10.1088/1361-6382/ab5c9a",
    journal = "Class. Quant. Grav.",
    volume = "37",
    number = "2",
    pages = "02LT01",
    year = "2020"
}

@article{Iacovelli:2022bbs,
    author = "Iacovelli, Francesco and Mancarella, Michele and Foffa, Stefano and Maggiore, Michele",
    title = "{Forecasting the Detection Capabilities of Third-generation Gravitational-wave Detectors Using GWFAST}",
    eprint = "2207.02771",
    archivePrefix = "arXiv",
    primaryClass = "gr-qc",
    doi = "10.3847/1538-4357/ac9cd4",
    journal = "Astrophys. J.",
    volume = "941",
    number = "2",
    pages = "208",
    year = "2022"
}

@article{Takeda:2019gwk,
    author = "Takeda, Hiroki and Nishizawa, Atsushi and Nagano, Koji and Michimura, Yuta and Komori, Kentaro and Ando, Masaki and Hayama, Kazuhiro",
    title = "{Prospects for gravitational-wave polarization tests from compact binary mergers with future ground-based detectors}",
    eprint = "1904.09989",
    archivePrefix = "arXiv",
    primaryClass = "gr-qc",
    doi = "10.1103/PhysRevD.100.042001",
    journal = "Phys. Rev. D",
    volume = "100",
    number = "4",
    pages = "042001",
    year = "2019"
}

@dataset{ImamLagosXGZenodo2026,
  author       = {Imam, Adil and Lagos, Macarena},
  title       = {XG-Scalar-Polarization-Forecast: Posterior Samples  for Bounds on Scalar Polarization with Next-Generation Gravitational-Wave Detectors},
  year         = {2026},
  publisher    = {Zenodo},
  doi          = {10.5281/zenodo.22291452},
  url          = {https://doi.org/10.5281/zenodo.22291452},
  note         = {[Dataset]}
}

@misc{ImamLagosXGCode2026,
  author  = {Imam, Adil and Lagos, M.},
  title   = {XG-Scalar-Polarization-Forecast: Analysis Code for Bounds on Scalar Polarization with Next-Generation Gravitational-Wave Detectors},
  year    = {2026},
  howpublished = {\url{https://github.com/adil1828/GW_JAX_TEAM_Nested/tree/main/XG_Scalar_Polarization_Forecast}},
  note    = {GitHub repository}
}

@article{DouchinHaensel2001,
  author  = {Douchin, F. and Haensel, P.},
  title   = {A unified equation of state of dense matter and neutron star structure},
  journal = {Astron. Astrophys.},
  volume  = {380},
  pages   = {151},
  year    = {2001},
  doi     = {10.1051/0004-6361:20011402},
  eprint  = {astro-ph/0111092},
  archivePrefix = {arXiv}
}

@article{Hinderer2010,
  author  = {Hinderer, Tanja and Lackey, Benjamin D. and Lang, Ryan N. and Read, Jocelyn S.},
  title   = {Tidal deformability of neutron stars with realistic equations of state and their gravitational wave signatures in binary inspiral},
  journal = {Phys. Rev. D},
  volume  = {81},
  pages   = {123016},
  year    = {2010},
  doi     = {10.1103/PhysRevD.81.123016},
  eprint  = {0911.3535},
  archivePrefix = {arXiv},
  primaryClass  = {astro-ph.HE}
}

@article{MullerSerot1996,
  author  = {M{\"u}ller, H. and Serot, B. D.},
  title   = {Relativistic mean-field theory and the high-density nuclear equation of state},
  journal = {Nucl. Phys. A},
  volume  = {606},
  pages   = {508},
  year    = {1996},
  eprint  = {nucl-th/9603037},
  archivePrefix = {arXiv}
}

@article{ReadEtAl2009,
  author  = {Read, Jocelyn S. and Lackey, Benjamin D. and Owen, Benjamin J. and Friedman, John L.},
  title   = {Constraints on a phenomenologically parametrized neutron-star equation of state},
  journal = {Phys. Rev. D},
  volume  = {79},
  pages   = {124032},
  year    = {2009},
  doi     = {10.1103/PhysRevD.79.124032},
  eprint  = {0812.2163},
  archivePrefix = {arXiv}
}

@article{Favata2014,
  author  = {Favata, Marc},
  title   = {Systematic Parameter Errors in Inspiraling Neutron Star Binaries},
  journal = {Phys. Rev. Lett.},
  volume  = {112},
  pages   = {101101},
  year    = {2014},
  doi     = {10.1103/PhysRevLett.112.101101},
  eprint  = {1310.8288},
  archivePrefix = {arXiv}
}

@article{WadeEtAl2014,
  author  = {Wade, Leslie and Creighton, Jolien D. E. and Ochsner, Evan and Lackey, Benjamin D. and Farr, Benjamin F. and Littenberg, Tyson B. and Raymond, Vivien},
  title   = {Systematic and statistical errors in a Bayesian approach to the estimation of the neutron-star equation of state using advanced gravitational-wave detectors},
  journal = {Phys. Rev. D},
  volume  = {89},
  pages   = {103012},
  year    = {2014},
  doi     = {10.1103/PhysRevD.89.103012},
  eprint  = {1402.5156},
  archivePrefix = {arXiv}
}

@article{Tolman1939,
  author  = {Tolman, R. C.},
  title   = {Static Solutions of Einstein's Field Equations for Spheres of Fluid},
  journal = {Phys. Rev.},
  volume  = {55},
  pages   = {364},
  year    = {1939},
  doi     = {10.1103/PhysRev.55.364}
}

@article{OppenheimerVolkoff1939,
  author  = {Oppenheimer, J. R. and Volkoff, G. M.},
  title   = {On Massive Neutron Cores},
  journal = {Phys. Rev.},
  volume  = {55},
  pages   = {374},
  year    = {1939},
  doi     = {10.1103/PhysRev.55.374}
}

@article{Tahura:2018zuq,
    author = "Tahura, Sharaban and Yagi, Kent",
    title = "{Parameterized Post-Einsteinian Gravitational Waveforms in Various Modified Theories of Gravity}",
    eprint = "1809.00259",
    archivePrefix = "arXiv",
    primaryClass = "gr-qc",
    doi = "10.1103/PhysRevD.98.084042",
    journal = "Phys. Rev. D",
    volume = "98",
    number = "8",
    pages = "084042",
    year = "2018",
    note = "[Erratum: Phys.Rev.D 101, 109902 (2020)]"
}

@article{Chamberlain:2017fjl,
    author = "Chamberlain, Katie and Yunes, Nicolas",
    title = "{Theoretical Physics Implications of Gravitational Wave Observation with Future Detectors}",
    eprint = "1704.08268",
    archivePrefix = "arXiv",
    primaryClass = "gr-qc",
    doi = "10.1103/PhysRevD.96.084039",
    journal = "Phys. Rev. D",
    volume = "96",
    number = "8",
    pages = "084039",
    year = "2017"
}

@article{Perkins:2020tra,
    author = "Perkins, Scott E. and Yunes, Nicol{\'a}s and Berti, Emanuele",
    title = "{Probing Fundamental Physics with Gravitational Waves: The Next Generation}",
    eprint = "2010.09010",
    archivePrefix = "arXiv",
    primaryClass = "gr-qc",
    doi = "10.1103/PhysRevD.103.044024",
    journal = "Phys. Rev. D",
    volume = "103",
    number = "4",
    pages = "044024",
    year = "2021"
}

@article{ET:2019dnz,
    author = "Maggiore, Michele and others",
    collaboration = "ET",
    title = "{Science Case for the Einstein Telescope}",
    eprint = "1912.02622",
    archivePrefix = "arXiv",
    primaryClass = "astro-ph.CO",
    doi = "10.1088/1475-7516/2020/03/050",
    journal = "JCAP",
    volume = "03",
    pages = "050",
    year = "2020"
}

@article{Agathos:2013upa,
    author = "Agathos, Michalis and Del Pozzo, Walter and Li, Tjonnie G. F. and Van Den Broeck, Chris and Veitch, John and Vitale, Salvatore",
    title = "{TIGER: A data analysis pipeline for testing the strong-field dynamics of general relativity with gravitational wave signals from coalescing compact binaries}",
    eprint = "1311.0420",
    archivePrefix = "arXiv",
    primaryClass = "gr-qc",
    doi = "10.1103/PhysRevD.89.082001",
    journal = "Phys. Rev. D",
    volume = "89",
    number = "8",
    pages = "082001",
    year = "2014"
}

@article{LIGOScientific:2021sio,
    author = "Abbott, R. and others",
    collaboration = "LIGO Scientific, VIRGO, KAGRA",
    title = "{Tests of General Relativity with GWTC-3}",
    eprint = "2112.06861",
    archivePrefix = "arXiv",
    primaryClass = "gr-qc",
    reportNumber = "LIGO-P2100275",
    doi = "10.1103/PhysRevD.112.084080",
    journal = "Phys. Rev. D",
    volume = "112",
    number = "8",
    pages = "084080",
    year = "2025"
}

@article{KAGRA:2013rdx,
    author = "Abbott, B. P. and others",
    collaboration = "KAGRA, LIGO Scientific, Virgo",
    title = "{Prospects for observing and localizing gravitational-wave transients with Advanced LIGO, Advanced Virgo and KAGRA}",
    eprint = "1304.0670",
    archivePrefix = "arXiv",
    primaryClass = "gr-qc",
    reportNumber = "LIGO-P1200087, VIR-0288A-12, JGW-P1808427",
    doi = "10.1007/s41114-020-00026-9",
    journal = "Living Rev. Rel.",
    volume = "19",
    pages = "1",
    year = "2016"
}

@article{LIGOScientific:2017vwq,
    author = "Abbott, B. P. and others",
    collaboration = "LIGO Scientific, Virgo",
    title = "{GW170817: Observation of Gravitational Waves from a Binary Neutron Star Inspiral}",
    eprint = "1710.05832",
    archivePrefix = "arXiv",
    primaryClass = "gr-qc",
    reportNumber = "LIGO-P170817",
    doi = "10.1103/PhysRevLett.119.161101",
    journal = "Phys. Rev. Lett.",
    volume = "119",
    number = "16",
    pages = "161101",
    year = "2017"
}

@article{Wu:2024yno,
    author = "Wu, Jie and Li, Jin",
    title = "{Prospects of constraining on the polarizations of gravitational waves from binary black holes using space- and ground-based detectors}",
    eprint = "2407.13590",
    archivePrefix = "arXiv",
    primaryClass = "gr-qc",
    doi = "10.1103/PhysRevD.110.084057",
    journal = "Phys. Rev. D",
    volume = "110",
    number = "8",
    pages = "084057",
    year = "2024"
}

\end{document}